\documentclass[12pt]{iopart}
\usepackage{graphicx}
\begin{document}
\title[Full-Wave Effects On Maxwellian Lower-Hybrid Wave Damping]{An Assessment Of Full-Wave Effects On Maxwellian Lower-Hybrid Wave Damping}

\author{S J Frank$^1$, J C Wright$^1$, 
I H Hutchinson$^1$, and P T Bonoli$^1$}

\address{$^1$ Plasma Science and Fusion Center, Massachusetts Institute of Technology, 
Cambridge MA, USA}

\begin{abstract}
Lower-hybrid current drive (LHCD) actuators are important components of modern day fusion experiments as well as proposed fusion reactors. However, simulations of LHCD often differ substantially from experimental results, and from each other, especially in the inferred power deposition profile shape. Here we investigate some possible causes of this discrepancy; ``full-wave'' effects such as interference and diffraction, which are omitted from standard raytracing simulations and the breakdown of the raytracing near reflections and caustics. We compare raytracing simulations to state-of-the-art full-wave simulations using matched hot-plasma dielectric tensors in realistic tokamak scenarios for the first time. We show that differences between full-wave simulations and raytracing in previous work were primarily due to numerical and physical inconsistencies in the simulations, and we demonstrate that good agreement between raytracing and converged full-wave simulations can be obtained in reactor relevant-scenarios with large ray caustics and in situations with weak damping.
\end{abstract}


\submitto{\PPCF}
\maketitle
\ioptwocol
\section{Introduction}
Lower-hybrid current drive (LHCD) is a method of tokamak heating and current drive in which slow waves are launched by a phased waveguide array \cite{Brambilla1976} with frequency $\omega$ in the lower-hybrid (LH) frequency range: $\Omega_i \ll \omega \ll \Omega_e$ where $\Omega_s = q_s B/m_s$ is the cyclotron frequency for species $s$ with subscripts $e$ and $i$ referring to electrons and ions respectively. The launched LH waves propagate into the tokamak and Landau damp on electrons (ELD). As the waves damp, they impart their energy and momentum to the plasma heating it and driving current. 

Lower-hybrid current drive has been demonstrated in many tokamak experiments \cite{Bernabei1982,Porkolab1984,Bartiromo1986,Moriyama1990,Jacquinot1991,Ide1992,Peysson2001,Wilson2009,Cesario2010,Wallace2011,Liu2015} and is a key component in the heating and current drive systems of some proposed fusion reactor designs \cite{Najmabadi2006, Sorbom2015}. Despite its widespread use, simulations of LHCD often predict different power deposition and current drive profiles than those measured in experiments\cite{Wallace2011, MumgaardThesis, Schmidt2011,Decker2014,Peysson2016,Yang2018,Garofalo2018}. This difficulty is attributable in part to the phenomenon known as the LH ``spectral gap". Lower-hybrid waves often damp despite having initial phase velocities much greater than their linear Maxwellian damping condition of $v_{ph,\parallel} = \omega/k_\parallel \sim 3v_{th,e}$ where $k_\parallel$ is the parallel component of the wave vector $\vec{k}$, and $v_{th,e}$ is the electron thermal speed. Waves launched with phase velocities $v_{ph,\parallel} \sim 6-8v_{th,e}$, are able to damp because a small portion of the launched LH wave spectrum experiences a phase-velocity downshift. The velocity downshifted portion of the wave spectrum fills the spectral gap in the Landau plateau providing high energy electrons for the high velocity component of the spectrum to damp on. Various mechanisms have been proposed as sources of the downshift that provides spectral gap closure including: geometric downshift of the wave velocity \cite{Bonoli1981,Bonoli1982,Bonoli1986}, wave interactions with turbulence \cite{Bonoli1981,Bonoli1982,Decker2014,Biswas2020,Biswas2021}, diffractional broadening of the wave spectrum \cite{Pereverzev1992,Wright2009,Wright2010,Shiraiwa2011,Wright2014}, and spectral broadening as a result of parametric wave interactions \cite{Cesario2004,Decker2014}. As each of these effects modify the wave spectrum differently, if the dominant effect or combination of effects is not properly included in a simulation then its results may differ substantially from the experimental measurements.

Another proposed cause of difficulties in LHCD prediction is the use of the raytracing approximation in simulations. Raytracing approximates the plasma Helmholtz equation:
\begin{equation}\label{eq:helmholtz}
    \nabla \times \nabla \times \vec{E} = \frac{\omega^2}{c^2}\left(\vec{E} + \frac{4\pi i}{\omega} \vec{J}_p \right),
\end{equation}
with the perturbed current $\vec{J}_p = \stackrel{\leftrightarrow}{\sigma} \cdot \vec{E}$ where $\stackrel{\leftrightarrow}{\sigma}$ is the plasma conductivity tensor. In the LH limit:
\begin{equation}
\frac{\Omega_i}{\omega} \ll 1 \ll \frac{\Omega_e}{\omega},    
\end{equation}
the hot-plasma dielectric tensor is \cite{BrambillaPlasmaWaves,WrightBertelli2014}:
\begin{equation}\label{eq:lhdiel}
\eqalign{
    \stackrel{\leftrightarrow}{\epsilon}_{LH} = \stackrel{\leftrightarrow}{\textrm{I}} + \frac{i\stackrel{\leftrightarrow}{\sigma}}{\epsilon_0 \omega} =\cr \left[\begin{array}{ccc}
\epsilon_\perp - \sigma N_\perp^2 & -i\epsilon_{xy} & 0 \\
i\epsilon_{xy} & \epsilon_\perp- (\sigma+\tau_{MP}) N_\perp^2 & iN_\parallel N_\perp \delta_\times \\
0 & -iN_\parallel N_\perp \delta_\times & \epsilon_\parallel
\end{array}\right],
}
\end{equation}
where,
\begin{equation}
    \epsilon_\perp = 1 + \frac{\omega_{pe}^2}{\Omega_e^2}  -\frac{\omega_{pi}^2}{\omega^2}
\end{equation}
\begin{equation}
    \epsilon_\parallel = 1 -  \frac{\omega_{pe}^2}{\omega^2} \zeta_{e}^2 \textrm{Z}^\prime(\zeta_{e}) -\frac{\omega_{pi}^2}{\omega^2}
\end{equation}
\begin{equation}
    \epsilon_{xy} = \frac{\omega_{pe}^2}{\omega\Omega_e}
\end{equation}
\begin{equation}
    \sigma = \frac{3}{2} \beta_i \frac{\Omega_i^2}{\omega^2} + \frac{3}{8}\beta_e \frac{\omega^2}{\Omega_e^2}.
\end{equation}
\begin{equation}
    \tau_{MP} \cong -\beta_e \zeta_{e} \textrm{Z}(\zeta_{e})
\end{equation}
\begin{equation}
    \delta_\times \cong -\beta_e \frac{\Omega_e}{\omega}\zeta_{e}^2\textrm{Z}^\prime(\zeta_{e})
\end{equation}
Here, \textrm{Z} is the plasma dispersion function \cite{FriedConte}, the refractive index $\vec{N} = \vec{k} c / \omega$, for species $s$ the plasma frequency $\omega_{ps} = \sqrt{q_s^2 n_s / \epsilon_0 m_s}$, the thermal velocity $v_{th,s} = \sqrt{2 T_s/m_s}$, $\zeta_s = \omega/k_\parallel v_{th,s}$, and $\beta_s = \frac{\omega_{ps}^2}{\Omega_s^2}\frac{v_{th,s}^2}{c^2}$. In conventional tokamaks $\beta_s \ll 1$, and to prevent parametric decay $\omega \ge 2\omega_{LH}$ \cite{Porkolab1977}. Because of this, the mode conversion term $\sigma$, and FLR damping terms $\delta_x$ and $\tau_{MP}$, are excluded from the dielectric as mode conversion of the LH wave only occurs near LH resonance \cite{Bonoli1984,Bonoli1985} and they are very small. In order to solve (\ref{eq:helmholtz}) using raytracing we take the $kL \gg 1$ limit, where $L$ is the plasma gradient scale length, and apply the WKB approximation. This process allows us to obtain the dispersion relation:
\begin{equation}
    D(\omega,\vec{k},\vec{x}) = P_4 N_\perp^4 + P_2 N_\perp^2 + P_0,
\end{equation}
where:
\begin{equation}
    P_4 = \epsilon_\perp
\end{equation}
\begin{equation}
    P_2 = (\epsilon_\perp + \epsilon_\parallel)(N_\parallel^2 - \epsilon_\perp) + \epsilon_{xy}^2
\end{equation}
\begin{equation}
    P_0 = \epsilon_\parallel[(N_\parallel^2 - \epsilon_\perp)^2 - \epsilon_{xy}^2],
\end{equation}
and the ray equations \cite{Bonoli1982}:   
\begin{equation}
    \frac{d\vec{x}}{dt} = -\frac{\partial D/\partial\vec{k}}{\partial D/\partial \omega}
\end{equation}
\begin{equation}
    \frac{d\vec{k}}{dt} = \frac{\partial D/\partial\vec{x}}{\partial D/\partial \omega}
\end{equation}
\begin{equation}
    \frac{dP_{ray}}{dt} = -2\gamma P_{ray}.
\end{equation}
Here $D$ is the plasma dispersion relation, $\vec{x}$ is the position vector, $P_{ray}$ is the power density associated with a ray, and $\gamma$ is the wave damping rate obtained from $\textrm{Im}[D]$ (using the method shown in \cite{Bonoli1984,Bonoli1985}). 

A criticism of raytracing is that in most tokamak experiments LH waves are `weakly damped' meaning LH waves are reflected many times from cutoffs as they propagate through the plasma prior to damping. At a cutoff, $\vec{k} \rightarrow 0$, invalidating the large $kL$ raytracing limit. The breakdown of raytracing around cutoffs is exacerbated by the short plasma scale lengths $L$ in the edge region where reflection from the $\omega > \omega_{pe}$ cutoff generally occurs. Further, there is the matter of ray caustics. A caustic is an internal tangency point where rays will focus and an internal reflection will occur causing the field calculated by raytracing to momentarily become singular. Cutoffs and caustics may be ameliorated using modified forms of raytracing that transform the k-vector to different set of coordinates \cite{Chapman2002,Lopez2020,Donnelly2021} or a ray kinetic equation \cite{Kupfer1993}. These techniques give uniformly valid solutions, however, they can be difficult to implement in real tokamak geometries and are not widely used for LHCD simulation. Perhaps most importantly, raytracing neglects the effects of diffraction and interference which may broaden the wave spectrum and modify wave damping rates (as the Landau damping rate $\propto |E_\parallel|^2$). 

Here we report major new developments of a ``full-wave" solver, TORLH, based on TORIC \cite{Brambilla1988,Brambilla1999,Wright2004} that solves the plasma Helmholtz equation in the LH limit, (\ref{eq:helmholtz}), directly in tokamak geometry. We use TORLH to perform simulations of Alcator C-Mod, DIII-D and EAST then compare the results to raytracing simulations performed using identical cases. Importantly, all of the simulations performed here, both raytracing and full-wave, utilize matched boundary conditions and the hot-plasma correction to the parallel dielectric term which can substantially affect wave trajectories \cite{WrightBertelli2014}. The TORLH and raytracing simulations here, for simplicity, employ Maxwellian electron and ion distribution functions. This avoids the complication of coupling the wave propagation and damping solution to a quasilinear diffusion and Fokker-Planck calculation needed to obtain a non-Maxwellian plasma response. Maxwellian simulations, in fact, should perhaps be more sensitive to spectral gap effects as the \textit{entire} launched spectrum must experience a phase-velocity downshift in order to damp. In non-Maxwellian simulations only a portion of the spectrum must downshift to fill in the Landau plateau, after this, non-Maxwellian damping enhances the damping of high phase velocity waves \cite{Bonoli1986}. 

\section{The TORLH Full-Wave Solver}\label{sec:torlh}

The TORLH full-wave solver uses a semi-spectral electric field discretization: 
\begin{equation}\label{eq:torlhdisc}
    \vec{E} = e^{i(n_\phi\phi - \omega t)}\sum_{m=-\infty}^{+\infty} \vec{E}^{(m)}(\psi)e^{im\theta}
\end{equation}
A spectral discretization is used in angular directions $\theta$ and $\phi$ with poloidal and toroidal mode numbers $m$ and $n_\phi$. A cubic Hermite interpolating polynomial finite element (FE) discretization is used in the radial direction $\psi$ to solve for the Fourier electric field coefficients $\vec{E}^{(m)}(\psi)$ on a flux surface. TORLH solves for a single toroidal mode $n_\phi$, but multiple simulations at different toroidal mode numbers may be superimposed to reconstruct a 3-D field \cite{Lee2014}. The use of a finite element basis in the radial direction, however, makes TORLH significantly faster ($\mathcal{N}$ fewer operations) and less memory intensive ($\mathcal{N}^2$ less memory) than full-spectral solvers such as AORSA \cite{Jaeger2002}, but the use of a Fourier basis along each flux surface in TORLH allows inclusion of hot plasma effects in a fully self-consistent manner as the Fourier basis properly accounts for non-locality in the dielectric response on the flux surface. The computational advantage in TORLH comes from the form of the matrix produced when discretizing the Helmholtz equation: TORLH produces a block tridiagonal matrix while AORSA produces a fully dense matrix. An important limitation of TORLH versus AORSA is in the definition of the wave vectors. The $k_\parallel$ at a given spatial location is precisely defined by the mode numbers; however the radial component of $k_\perp$ ($k_\psi$) is represented by the finite elements and cannot be precisely defined (AORSA's fully spectral basis exactly defines both, but as the Maxwellian hot plasma LH wave dispersion relation may be written independently of $k_\perp$ this difference is generally unimportant in the TORLH simulations here):
\begin{equation}\label{eq:kpartorlh}
     k_\parallel = m (\hat{b}\cdot\nabla\theta) + n_\phi (\hat{b}\cdot\nabla\varphi)
\end{equation}

\begin{equation}
    \vec{k}_\perp = \vec{k}_\eta + \vec{k}_\psi
\end{equation}

\begin{equation}\label{eq:ketatorlh}
    k_\eta = m (\hat{b}\cdot\nabla\varphi) - n_\phi (\hat{b}\cdot\nabla\theta)
\end{equation}
where $\hat{b} = \vec{B_0}/|B_0|$ is the background magnetic field vector. Discretization (\ref{eq:torlhdisc}) is used to solve the hot-plasma Helmholtz equation in the LH limit:
\begin{equation}\label{eq:torlhhelmholtz}
    \nabla \times \nabla \times \vec{E} = \epsilon_\perp\vec{E}_\perp + i\epsilon_{xy}(\hat{b}\times\vec{E}_\perp) + \epsilon_\parallel E_\parallel \hat{b}. \,
\end{equation}
In TORLH, Helmholtz's equation (\ref{eq:torlhhelmholtz}) is put in Galerkin weak form and its dimension is reduced removing the radial field component $E_\psi$ (this reduction may be performed as we have ordered out the term $\propto \sigma N_\perp^2$ corresponding to the pressure driven wave that occurs about the LH resonance and does not appear in LHCD scenarios \cite{Wright2009}). Using discretization (\ref{eq:torlhdisc}) a block tridiagonal matrix is produced which may be inverted to produce an electric field solution using a custom 3-D parallelized block-cyclic reduction solver \cite{Lee2014}.

Because of its specialized discretization, highly parallelized solver, and recent upgrades to its post-processing algorithm made during this work, TORLH is capable of performing converged simulations of LH wave electric fields in nearly all present day LHCD experiments using available supercomputers, making it one of the only tools able to perform direct verification of raytracing and assessment of full-wave effects in LHCD experiments. We will now discuss the convergence requirements in TORLH and the implementation of a modified waveguide boundary condition required to ensure good agreement with raytracing.

\subsection{Convergence in TORLH}

Both the spectral and FE convergence requirements of TORLH are dictated by the LH accessibility and electron Landau damping limits \cite{Bonoli1985}:
\begin{equation}\label{eq:lhaccess}
    \frac{\omega_{pe}}{\Omega_e} + \sqrt{1+\frac{\omega_{pe}^2}{\Omega_e^2}-\frac{\omega_{pi}^2}{\omega^2}} \leq N_\parallel < \frac{c}{3v_{the}}    .
\end{equation}
Here the lower limit is set by mode conversion to the fast wave and the upper limit is set by Landau damping. Combining the accessibility limits with the cold plasma electrostatic dispersion relation:
\begin{equation}\label{eq:lhestat}
    k_\perp^2 = -\frac{P}{S}k_\parallel^2 \sim \frac{\omega_{pe}^2}{\omega^2}k_\parallel^2
\end{equation}
allows us to estimate the $k_\perp$ at each accessibility limit. 

\begin{figure}
    \centering
    \includegraphics[width=0.5\textwidth]{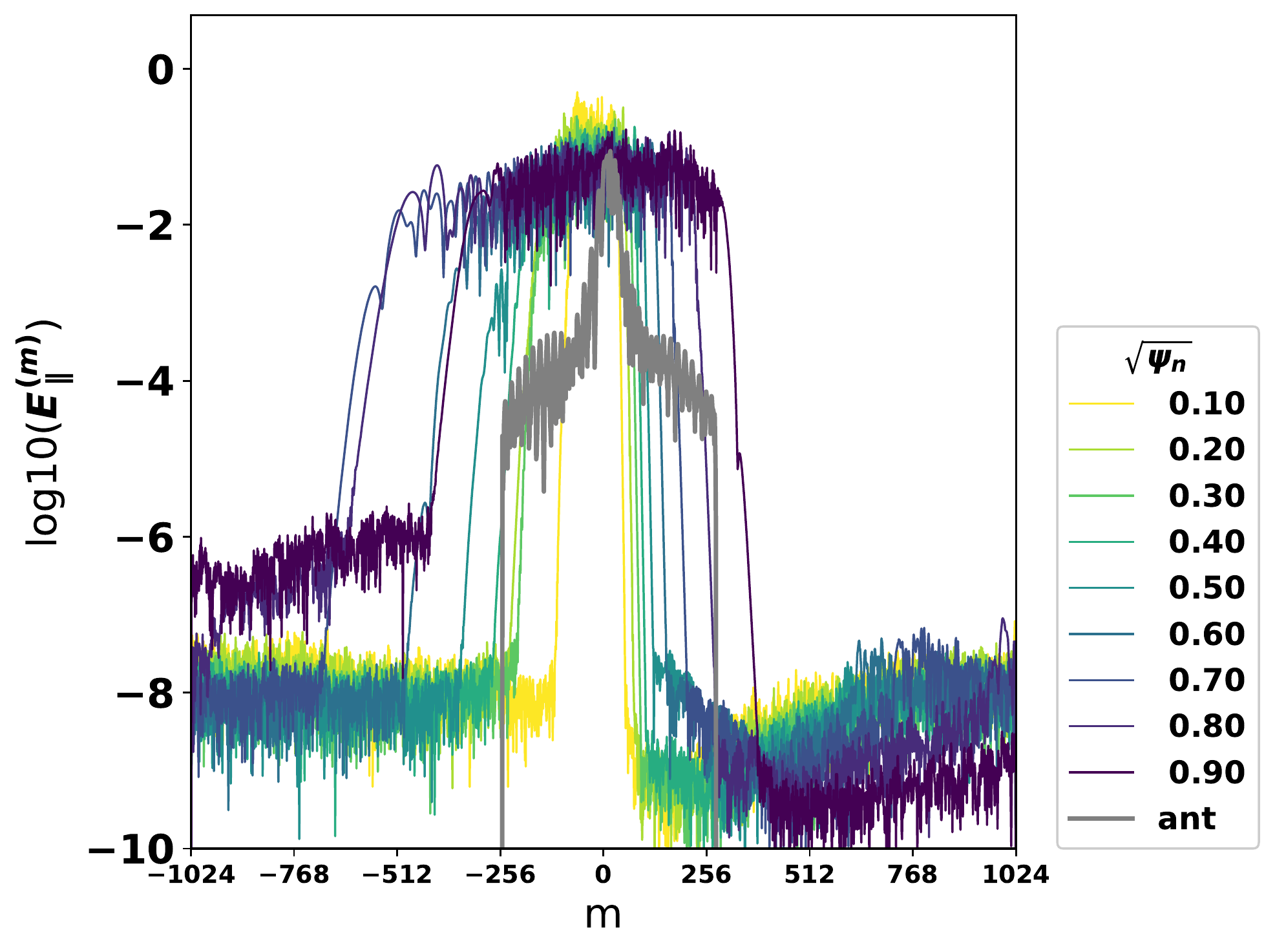}
    \caption{Spectral convergence plots showing Fourier coefficients $E^{(m)}$ versus mode number $m$ on labeled flux surfaces in normalized square root poloidal flux $\sqrt{\psi_n}$ for an Alcator C-Mod simulation with launch $N_\parallel = -1.6$. The asymmetry observable the mode spectrum here is a result of the lower and upper accessibility limits described in Eq. (\ref{eq:lhaccess}). Positive mode numbers continue until they reach the mode-conversion limit and negative mode numbers are eventually limited by Landau damping.}
    \label{fig:spec_conv}
\end{figure}

\subsubsection{Spectral Convergence}

Spectral convergence is easily evaluated in full-wave simulation codes by analyzing the relative magnitude of the Fourier coefficients in the mode spectrum. In the case of TORLH the condition for good convergence of the Fourier basis is well established and corresponds to the number of modes needed to resolve the largest $k_\perp \sim k_\eta \sim m/r$ in the system. Using (\ref{eq:lhestat}) and (\ref{eq:lhaccess}) this convergence requirement may be written:
\begin{equation} \label{eq:specconv1}
    m \cong \frac{\omega_{pe,a}}{3v_{the,0}}a
\end{equation}
\begin{equation}\label{eq:specconv2}
    \mathcal{N}_m = 2m + 1,
\end{equation}
where $\mathcal{N}_m$ is the number of modes needed in a TORLH simulation. The Landau damping limit is used to formulate the convergence criterion here as it almost always requires a larger mode number deviation from the launched $m=0$ mode to resolve. Despite being approximate, (\ref{eq:specconv1}) provides remarkably accurate estimations of the spectral convergence requirement. For the example C-Mod case shown in Figure~\ref{fig:spec_conv}, we see that for a launch $N_\parallel = -1.6$ using $\mathcal{N}_m = 2047$ spectral convergence is obtained. In this case $n_{e,a} \sim 2.5 \times 10^{19}$ m$^{-3}$ (this is the approximate density at $\sqrt{\psi_n} = 0.95$), $T_{e0} = 2.33$ keV, and $a=0.22$ m. Equation (\ref{eq:specconv2}) predicts here that $\mathcal{N}_m \sim 1620$. This prediction is remarkably close to the actual requirement.

\begin{figure*}
    \centering
    \includegraphics[width = 0.48\textwidth]{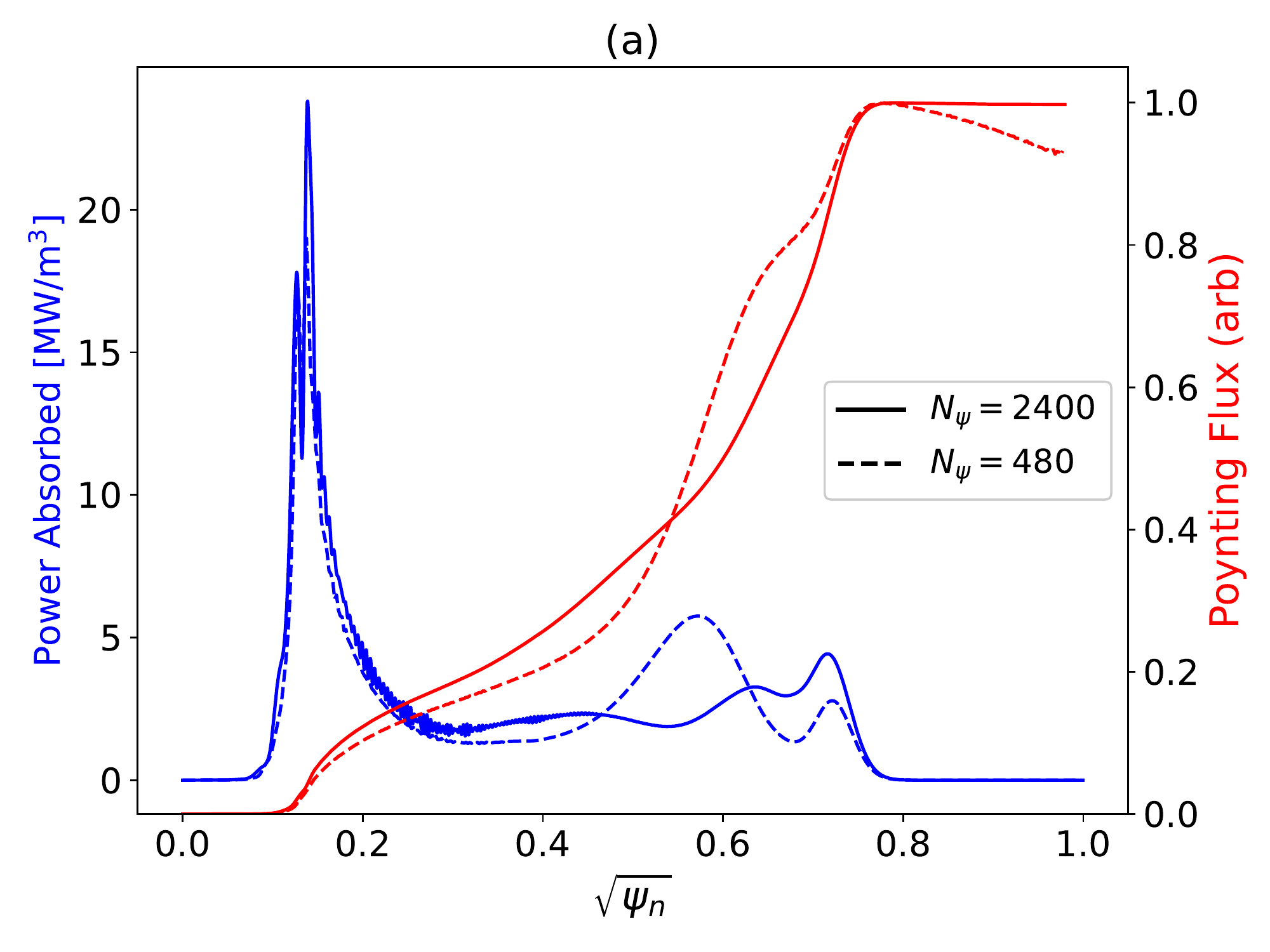}
    \includegraphics[width = 0.48\textwidth]{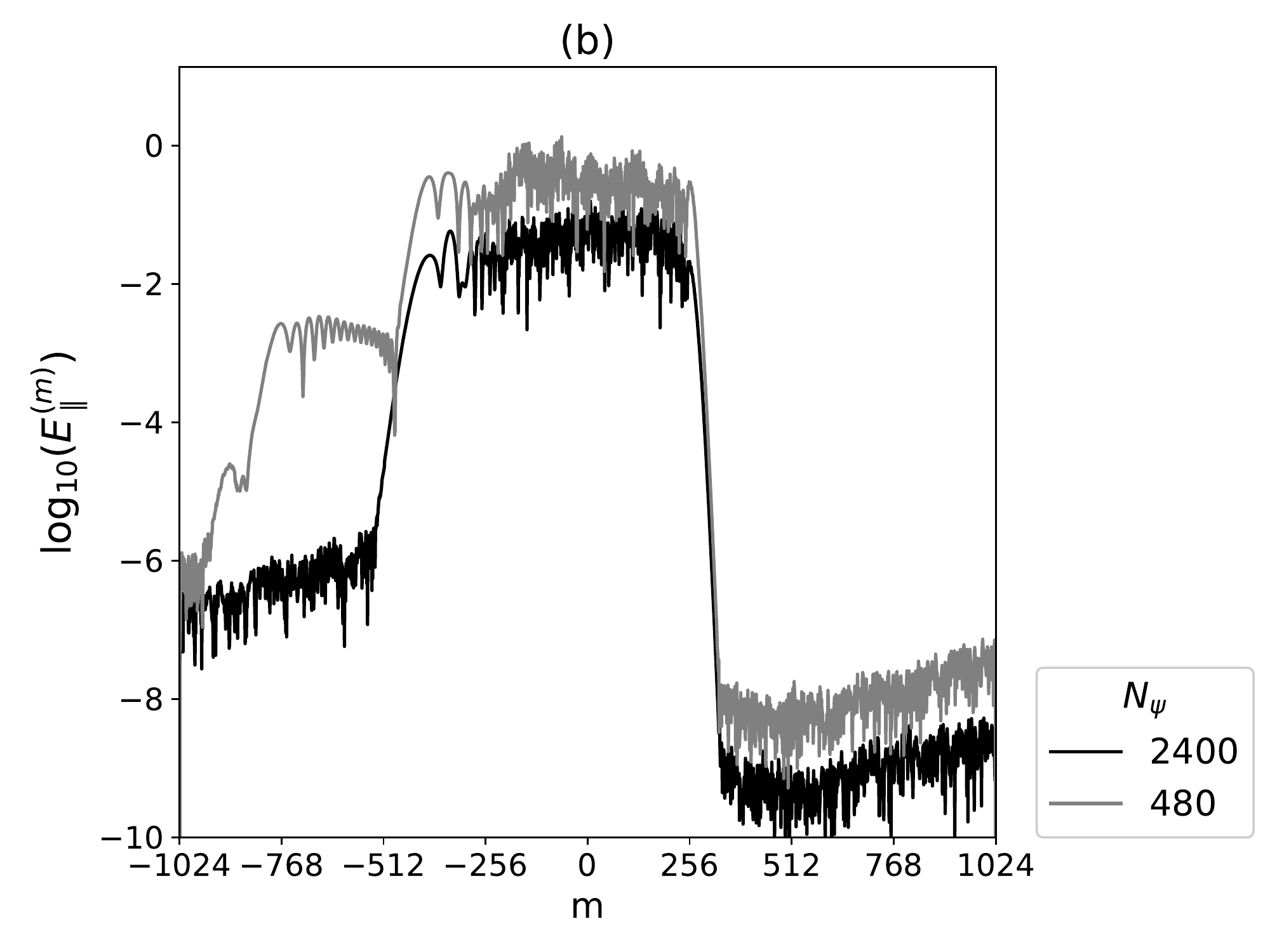}
    \caption{(a) Plots of the normalized Poynting flux and power deposition (normalized to 1MW of launch power) calculated by TORLH in an Alcator C-Mod discharge with a launched $N_\parallel = -1.6$ for two different values of $\mathcal{N}_\psi$. As convergence of the FE basis improves the droop in the Poynting vector at the edge disappears and the power deposition profile moves on axis. (b) Plots of the Fourier coefficients $E^{(m)}$ versus poloidal mode number $m$ at $\sqrt{\psi_n} = 0.85$ for two different values of $\mathcal{N}_\psi$. As the FE number is increased the magnitude of high $m$ modes drops exponentially.}
    \label{fig:fe_conv}
\end{figure*}

\subsubsection{Finite Element Convergence}

While the Fourier mode convergence requirements of TORLH are well documented \cite{Yang2018,Wright2014,Wright2010,Wright2009}, the finite element convergence requirements have not received similar attention. Typically, it had been assumed that one finite element per perpendicular wavelength, $\lambda_\perp = 2\pi/k_\perp$, was sufficient to resolve TORLH simulations. However, here we have found that the condition is in fact substantially more stringent. In order to establish a FE convergence requirement we simulated the C-Mod discharges originally investigated by Schmidt \cite{SchmidtThesis} and previously simulated with TORLH \cite{Wright2009,Wright2014} while varying the number of finite elements from 480-4800 elements or 1-10 elements per $\lambda_{\perp,min}$. C-Mod simulations were preferred here as they converge at smaller simulation scales than larger tokamaks. Convergence was assessed by analyzing the Poynting flux and the poloidal mode spectrum, as illustrated by the results of the C-Mod FE scan in Figure~\ref{fig:fe_conv}. Poor convergence of the finite element basis causes a characteristic droop, and in cases of exceptionally poor convergence, dramatic oscillation of the Poynting flux at the edge of the plasma in addition to an increase in $E^{(m)}$ at large $m$ number. When underresolved, waves with a large $k_\parallel$ have evanescence lengths that are too short to be resolved with the finite element basis \cite{TORICManual}. This causes a growing numerical mode and spectral pollution which is especially evident in the edge. Growing modes when the FE resolution is too low cause a characteristic increase in the $E_\parallel^{(m)}$ values at high poloidal mode number and oscillation in the Poynting flux. These modes will not damp until they experience a $k_\parallel$ downshift allowing them to once again be resolved and damp farther inside the core. FE resolution scans found that there is a FE requirement of roughly 5-10 FE per shortest perpendicular wavelength to suppress the pollution phenomenon and provide converged power deposition and Poynting flux profiles. Expressed similarly to the spectral convergence requirement:
\begin{equation}\label{eq:feconv}
    \mathcal{N}_\psi \cong \frac{5a\kappa k_{\perp,max}}{2\pi} \sim \frac{5a \kappa \omega_{pe,a}}{6\pi v_{the,a}},
\end{equation}
where $\kappa$ is the plasma elongation and 5 FE per wavelength was assumed. Using the parameters from the Schmidt C-Mod experiments \cite{Schmidt2011} once again, (\ref{eq:feconv}) predicts that $\mathcal{N}_\psi \sim 2400$ is needed to produce a converged solution. This is borne out well in simulation as shown in Figure~\ref{fig:fe_conv} where it is shown the Poynting flux exhibits monotonic behavior for $\mathcal{N}_\psi = 2400$. It was also found power deposition profiles remained constant for $\mathcal{N}_\psi \ge 2400$. This is a much larger value of $\mathcal{N}_\psi$ than that used in prior studies \cite{Wright2014,Wright2010,Wright2009, Yang2018, Yang2014}, and as FE convergence can have a non-negligible effect on power deposition, using sufficient $\mathcal{N}_\psi$ is important to effectively compare full-wave simulations versus raytracing. In cases where plasmas are strongly shaped or there is a large pedestal, such as the DIII-D simulations performed in Section~\ref{sec:sims_diiid}, even larger finite element numbers than those specified by (\ref{eq:feconv}) can be required to resolve rapid poloidal Jacobian variations near the plasma edge. 

Satisfying the much larger finite element convergence requirement necessitated major upgrades to the parallel post-processing algorithm in TORLH that calculates the power deposition in the plasma after solving for the electric fields. Previously, simulations would run out of memory during the post-processing step and finite element number was limited to approximately two-thousand elements for a system with 4 GB of memory per MPI rank. After aggressive memory management was implemented and some calculations were rewritten using more computationally efficient Fourier-space formulations, the memory burden of this portion of the code was reduced by over an order of magnitude, alleviating finite element limitations.  

\begin{figure}
    \centering
    \includegraphics[width = 0.48\textwidth]{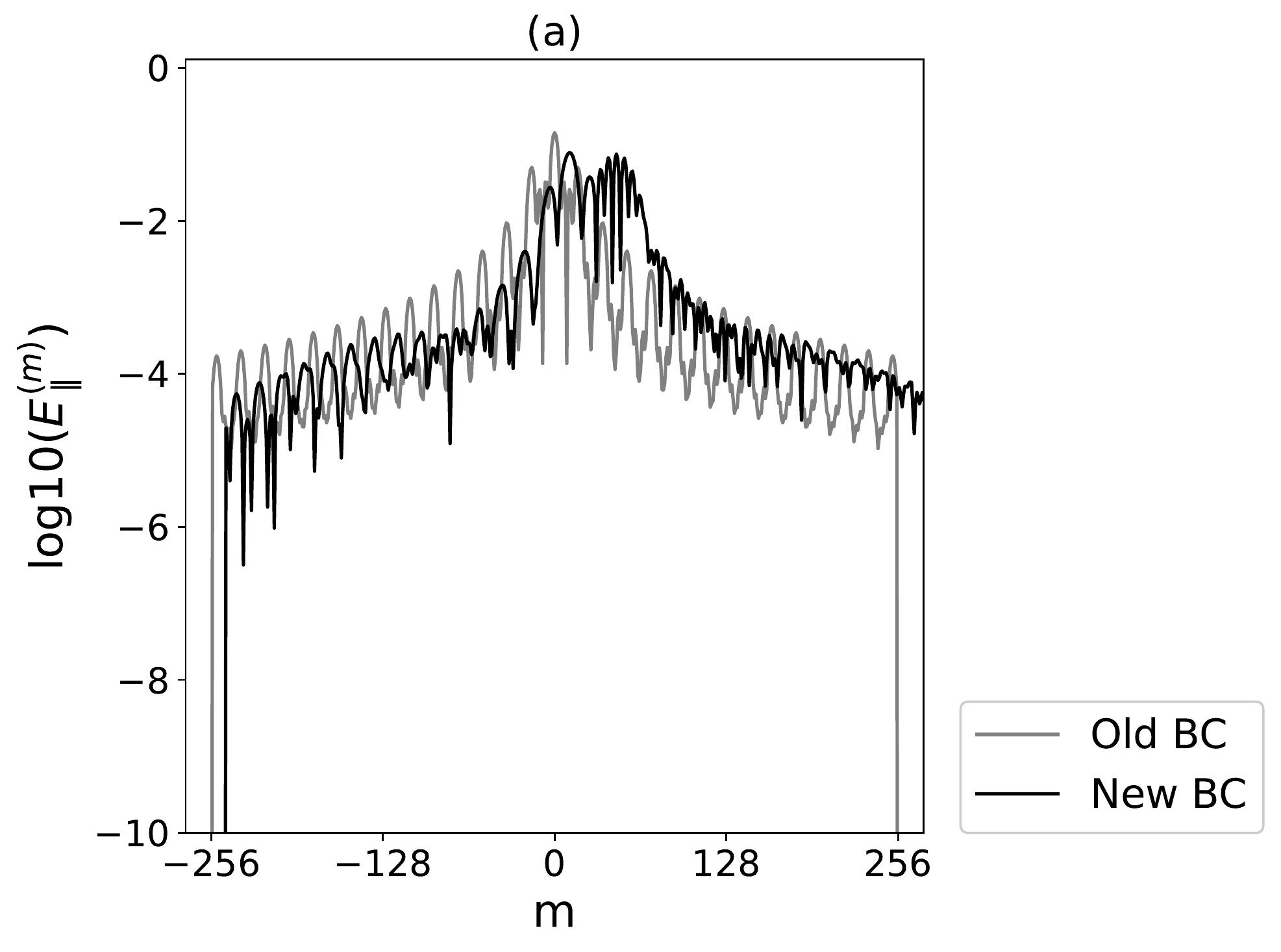}
    \includegraphics[width = 0.48\textwidth]{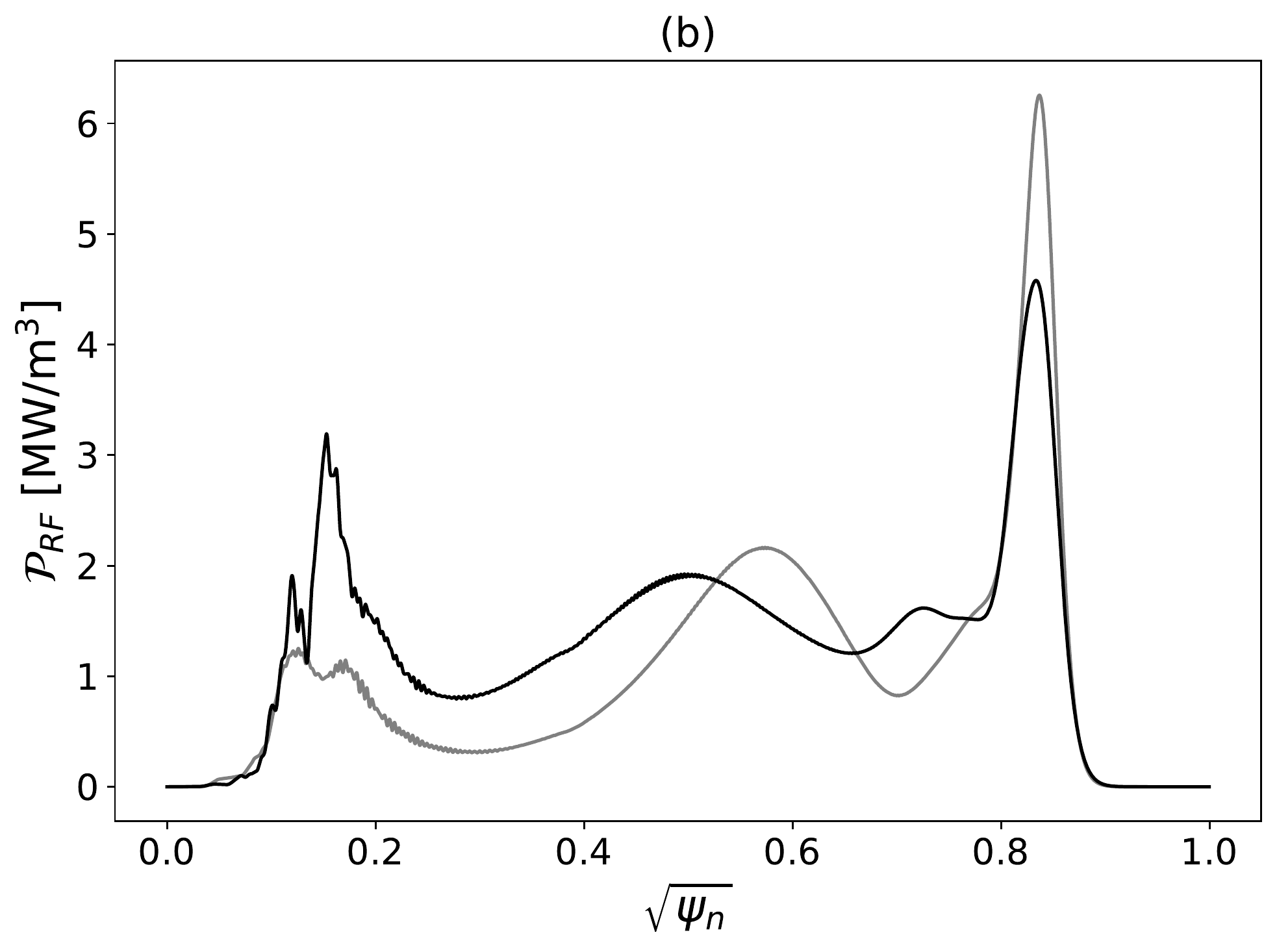}
    \caption{Comparisons of the old boundary condition, grey lines, and new boundary condition black lines, for: (a) Spectral plots showing Fourier coefficients $E^{(m)}$ versus mode number $m$ for the corrected and uncorrected antenna boundary condition in Alcator C-Mod. The C-Mod launcher uses four rows of waveguides located at $\theta_g = \pm 10 \,$ and $\, \pm 30$ degrees off of the outboard midplane, and in these simulations were launching a $n_\parallel = -3.1$. (b) Power deposition profiles from the corrected and uncorrected antenna boundary conditions. A noticeable change in peak heights and some peak locations results from correcting the boundary condition.}
    \label{fig:BC_corr}
\end{figure}

\subsection{The Improved TORLH Boundary Condition}\label{sec:torlh_bc}

An important modification to the TORLH boundary condition was made over the course of this work in order to enable more accurate replication of raytracing simulations. TORLH uses a fundamental boundary condition to replicate a waveguide grill \cite{Brambilla1976} by exciting a parallel electric field at the simulation boundary. Letting $\theta_g$ be the center of the waveguide grill the excited field at the waveguide mouth, $\psi_w$ corresponding to the edge of the simulation domain is:
\begin{equation}\label{eq:BCreal}
    E_\parallel(\psi_w, \theta) = E_0 \cos{\left(\pi \frac{\theta - \theta_g}{2\Delta_g}\right)},
\end{equation}
within $\Delta g$ of the waveguide mouth and zero elsewhere. 
The amplitude $E_0$ is typically set to a normalized value of 1 V/m in TORLH, $\Delta_g = h/2N_\tau$ with $N_\tau$ an equilibrium magnetic field metric coefficient which is $\propto a$ representing the effective minor radius. Fourier transforming (\ref{eq:BCreal}) yields:
\begin{equation}\label{eq:BCfourier}
    \eqalign{E_\parallel^{(m)} (\psi_w,m) \cr= \sum_n \frac{2\pi}{\pi^2-4m^2\Delta_{g,n}^2}\cos{(m\Delta_{g,n})}e^{-im\theta_{g,n}}},
\end{equation}
where we have written the boundary condition in terms of the Fourier coefficients imposed at the edge of the simulation domain for $n$ waveguides.

This boundary condition works well when there is only a single waveguide centered at $\theta_g$. However, when there are multiple waveguides, as is often the case in TORLH, or the waveguide's poloidal arc length becomes large this BC no longer will launch effectively a single $N_\parallel$ for fixed toroidal mode number $n_\phi$. Instead, a spectrum will be launched with $N_{\parallel} \sim n_\phi c R_{grill}/\omega$, where $R_{grill}$ is the radial location of each launcher grill in real-space. In order to approximate the launch of a single $N_\parallel$ value one must add an offset to the value of $m$, i.e. $m=m+m_{off}$, in (\ref{eq:BCfourier}) based on the radial location of each waveguide. This correction may be compactly written $m_{off} = -[k_\parallel - n_\phi (\vec{b}\cdot\nabla\phi)]/(\vec{b}\cdot\nabla\theta)$, and obtained by solving (\ref{eq:kpartorlh}) for m with $k_\parallel$ corresponding to the desired launch $N_\parallel$. In reality, an LHCD waveguide launches a spectrum of $N_\parallel$ values rather than the single value of $N_\parallel$ we have modeled in TORLH. However, the primary $n_\phi$ which is launched varies poloidally over the launcher resulting in a quantitatively different $N_\parallel$ spectrum than the TORLH boundary condition. Furthermore, in order to perform closely matched simulations with GENRAY it is much easier to launch a single $n_\phi$ and $N_\parallel$ in TORLH. This allows the us to use a single value of $N_\parallel$ without further spectrum matching for the GENRAY initial conditions. An example of an offset corrected boundary condition versus a boundary condition without an offset for the 4-row waveguide used in Alcator C-Mod is shown in Figure~\ref{fig:BC_corr}. Matching of the boundary condition was of great importance to achieving consistent agreement with the raytracing simulations performed here, because small discrepancies in LH waves' initial launch location and spectrum can greatly affect their propagation \cite{Bonoli1982}. The mismatch between the intended launch $N_\parallel$ without this correction could, in situations of $2-3$ pass damping such as those analyzed previously in \cite{Wright2009}, cause substantial disagreement with raytracing results.

\section{Comparisons w/ Raytracing}\label{sec:sims}

In order to analyze the impact of full-wave effects and the breakdown of the WKB approximation on the validity of raytracing, we have prepared a set of tightly matched test cases in which Maxwellian raytracing and full-wave solutions to LH wave propagation and damping may be compared. These cases used TORLH and a modified version of GENRAY \cite{Smirnov1994} using the same version of the hot-plasma dielectric in the LH limit as TORLH (\ref{eq:lhdiel}), including the hot plasma correction to the real part of the parallel dielectric term that has an important effect on LH wave trajectories, first noted in \cite{WrightBertelli2014}. Each case was prepared using the Integrated Plasma Simulator (IPS) \cite{Elwasif2010} and both TORLH and GENRAY were initialized from identical plasma state files (these include equilibrium and plasma profile information). The two codes used identical conducting wall boundary conditions and launch spectra. Neither GENRAY or TORLH included collisional wave damping. Here, we only simulated the tokamak core and did not include a realistic scrape-off layer model. Because of this, the power absorbed by collisional damping was very small, P$_{coll} \sim$ 5\% P$_{LD}$. In addition to verification of raytracing in the tokamak core, these test-cases serve to demonstrate the recent upgrades to the TORLH solver and advances in supercomputing which allow us to run TORLH at extremely large scales (tens of thousands of processors). Here converged simulations of moderately sized tokamak experiments such as DIII-D and EAST are performed which were previously not possible.  

\begin{figure*}
    \centering
    \includegraphics[width=\textwidth]{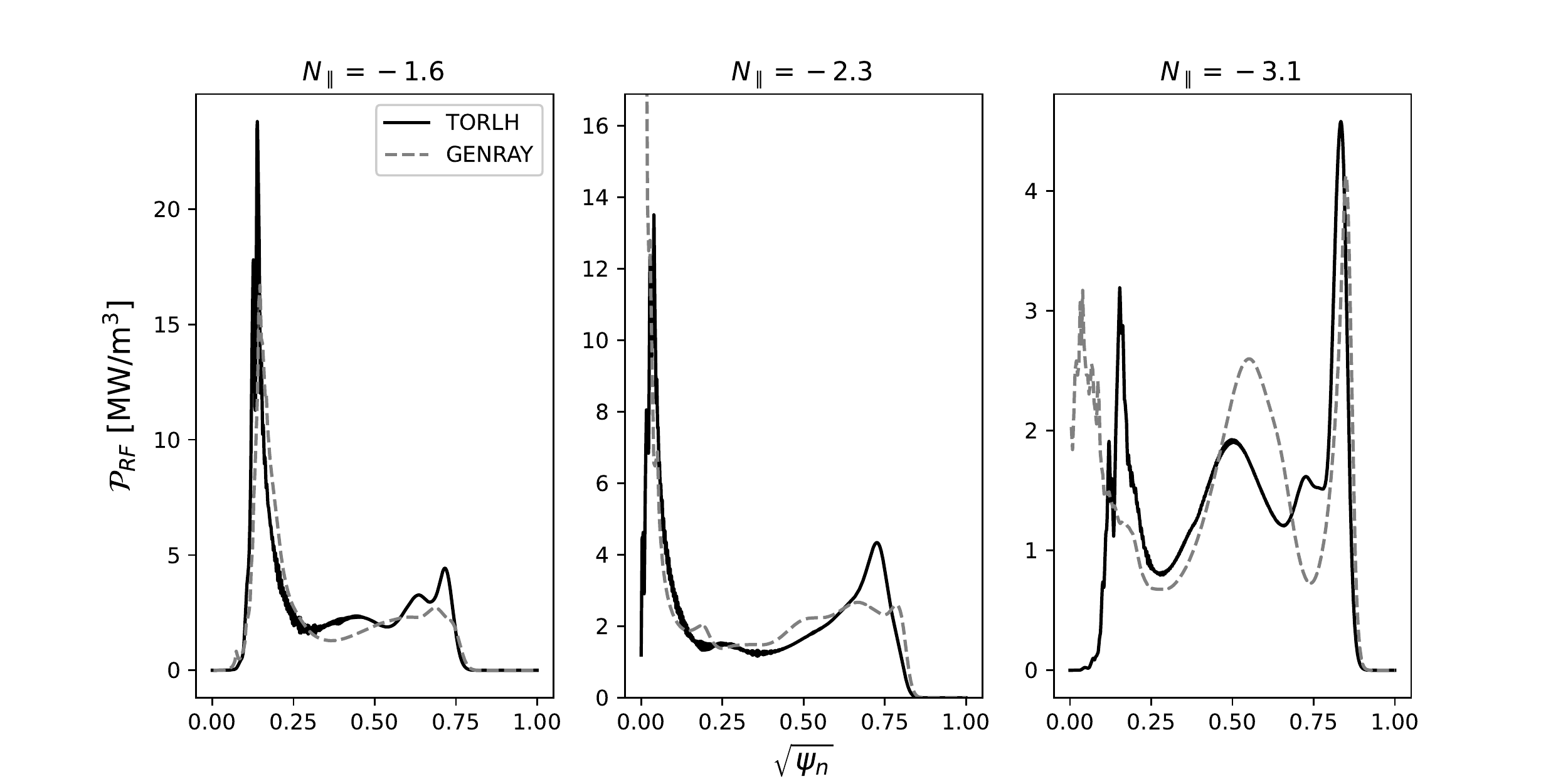}
    \caption{Absorbed power in MW/m$^3$ versus square root poloidal flux $\sqrt{\psi_n}$ for 1 MW of launched power in Alcator C-Mod at three different values of $N_\parallel$ calculated by TORLH (solid) and GENRAY (dashed). Good agreement is obtained at each value of $N_\parallel$ with GENRAY accurately reproducing the power deposition profiles calculated in TORLH across damping regimes. Differences which exist may be easily attributed to minor computational differences between the two codes which could not easily be eliminated. For example, the on-axis discrepancies where $\sqrt{\psi_n} < 0.05$ are due to differences in how the two codes handle interpolation and volume binning near the magnetic axis.}
    \label{fig:cmod_scan}
\end{figure*}

\subsection{C-Mod $N_\parallel$ Scan}\label{sec:sims_cmod}

The first simulations we performed were a replication of the $N_\parallel$ scan performed by Wright \cite{Wright2009, Wright2010,Wright2014} on Alcator C-Mod shot \#1060728011. While this tokamak discharge has been the subject of many previous raytracing and full-wave simulations \cite{SchmidtThesis,Wright2009,Wright2010,Shiraiwa2011,Wright2014}, no previous study applied a hot plasma correction to the raytracing results, imposed strict matching between simulations, and ensured total convergence of the full-wave simulations. The simulations here used three different values of $N_\parallel$: -1.6, -2.3, and -3.1, corresponding to the 60, 90, and 120 degree launcher phasings. In all simulations the LHCD launcher was modeled by four waveguide grills placed at $\theta =$ -30, -10, 10, \& 30 degrees relative to the outboard mid-plane (the same launcher configuration used in Wright \cite{Wright2009,Wright2010,Wright2014}), and great care was taken to ensure that the waveguides in TORLH and GENRAY were precisely aligned. GENRAY simulations used 100 rays equi-spaced poloidally along each waveguide for a total of 400 rays, and all rays were launched at a single $N_\parallel$ value corresponding to the launched $N_\parallel$ specified in TORLH. All TORLH simulations used resolutions of $\mathcal{N}_m = 2047$ and $\mathcal{N}_\psi =2400$ and were run on 8128 processors on Cori at NERSC for approximately 20 minutes each (The short runtime and large processor count is due to the good performance scaling of the matrix inversion algorithm and the memory constraints \cite{Lee2014}. The matrix to be inverted in TORLH must be distributed and stored in the RAM throughout the inversion). The plasma parameters and profiles used here were identical to previous modeling studies \cite{Wright2009,SchmidtThesis} with $T_{e0} = 2.33$ keV, $n_{e0} = 7.0 \times 10^{19}$ m$^{-3}$, and $B_0 = 5.4$ T and a magnetic equilibrium from EFIT \cite{Lao1985}. Both GENRAY \& TORLH simulations were initialized with the IPS using identical plasma states to further ensure the simulations were completely consistent. 

The results of these simulations are shown in Figure~\ref{fig:cmod_scan}. In all cases good agreement between raytracing and full-wave simulation was obtained; both power deposition profiles as well as the ray trajectories and full-wave field patterns closely matched. This demonstrates that raytracing can accurately reproduce full-wave simulation results across damping regimes. In the $N_\parallel=-1.6$ and $-2.3$ cases the LH wave was weakly damped, while in the $N_\parallel = -3.1$ case the wave was damped much more strongly in 2-3 passes. These results notably differ from previous analysis of Maxwellian damping in this discharge \cite{Wright2009} where poor agreement was obtained between GENRAY and TORLH as $N_\parallel$ increased. This disagreement was attributed to the formation of caustic surfaces in the raytracing simulations but in fact was due to a combination of: mismatched boundary/initial conditions, failure to include hot-plasma corrections to wave propagation in the raytracing simulations, and insufficient radial resolution in the TORLH simulations. These results demonstrate that for a low density, moderate aspect ratio $\epsilon = a/R \sim 0.3$, Maxwellian plasma across damping regimes, raytracing accurately simulates the core propagation and damping of LH waves and full-wave effects appear to be of little importance. The remaining differences which exist in the modeled power deposition profiles may be plausibly explained by boundary condition and equilibrium mismatches which could not easily be eliminated between the two codes, and the influence of some diffractional broadening in the full-wave simulations \cite{Pereverzev1992}. This is indicated by the fact that as $N_\parallel$ is decreased the profiles' agreement becomes more precise. As damping becomes weak the lower-hybrid wave quickly becomes stochastic in high toroidicity plasmas. In these cases toroidal broadening of the wave spectrum dominates, and differences due to diffraction as well as boundary and initial conditions become relatively unimportant. These full-wave results run contrary to previous approximate results that postulated diffractional broadening would be of greater importance than toroidally induced broadening \cite{Pereverzev1992}.    

\begin{figure}
    \centering
    \includegraphics[width=0.5\textwidth]{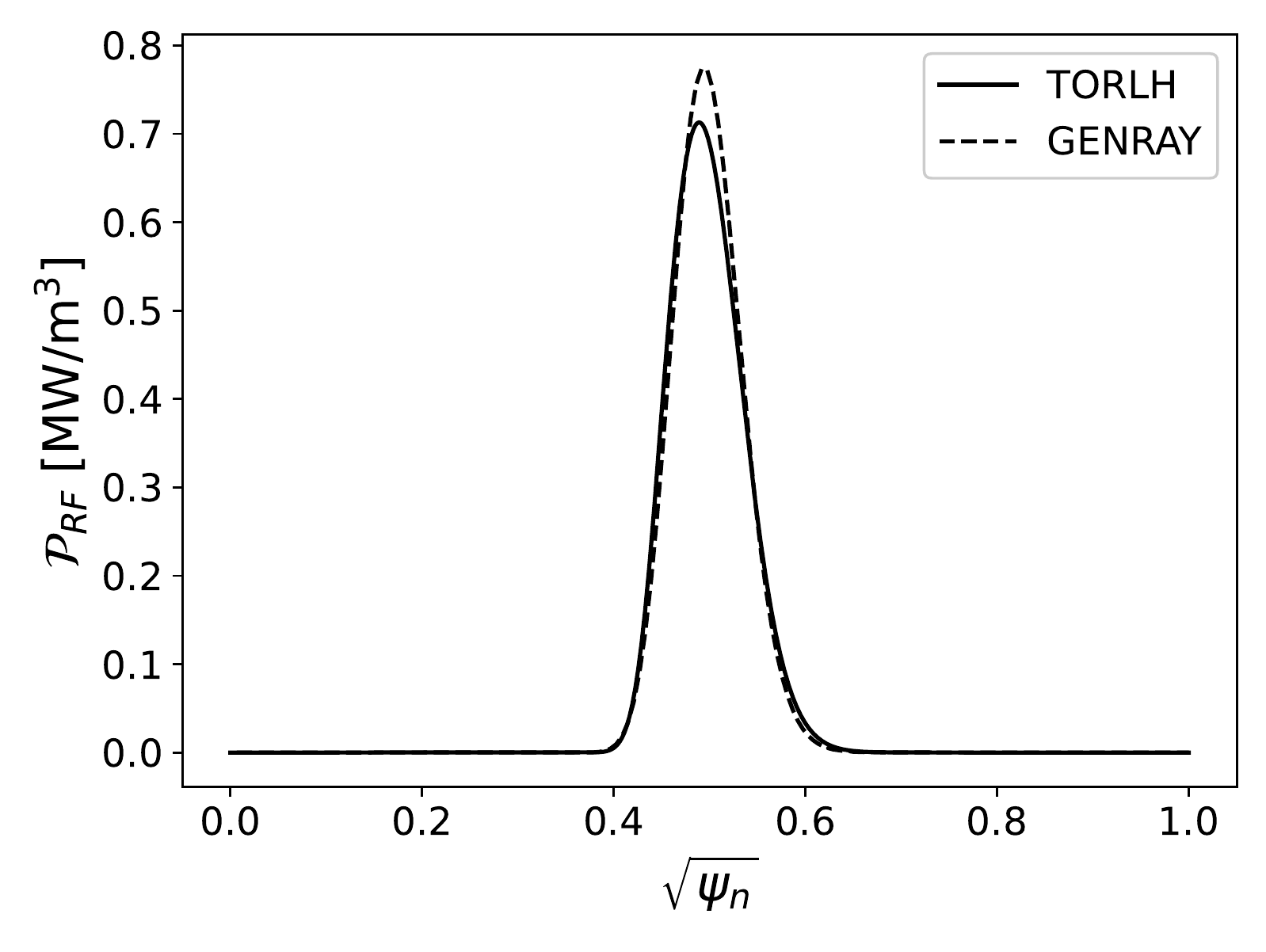}
    \includegraphics[width=0.5\textwidth]{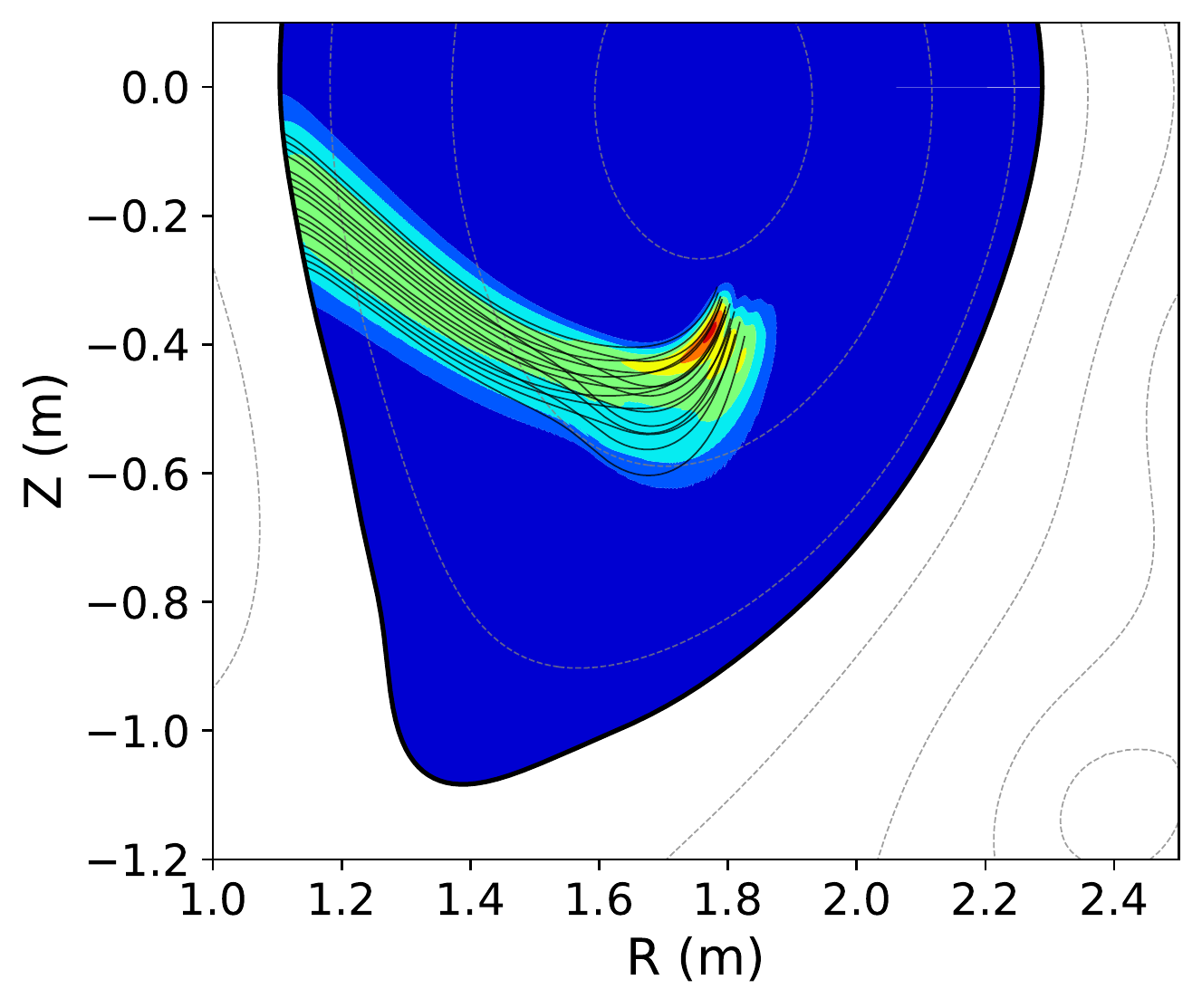}
    \caption{(a) Absorbed power in MW/m$^3$ versus square root poloidal flux $\sqrt{\psi_n}$ for 1 MW of launched power in DIII-D using the high field side launcher. (b) GENRAY rays and TORLH fields plotted over the DIII-D discharge \#174658 equilibrium.}
    \label{fig:diiid}
\end{figure}

\subsection{DIII-D HFS LHCD}\label{sec:sims_diiid}

In this test case we replicate a discharge from the DIII-D high-field side (HFS) LHCD launcher design study \cite{Seltzman2019}. We used T,n profiles and an equilibrium from DIII-D shot \#174658: a high performance, non-inductive, QH-mode discharge \cite{Burrell2020}. The DIII-D HFS LHCD launcher experiment utilizes the improved wave accessibility on the HFS to achieve efficient LHCD in reactor relevant high-$\beta$ advanced tokamak plasmas. HFS wave accessibility is improved because the linear mode conversion cutoff in the LHCD accessbility condition (\ref{eq:lhaccess}), goes as $N_{\parallel,min} \propto \frac{\sqrt{n_e}}{B}$. Therefore, launching on the HFS gives access to higher phase velocity waves which are inaccessible on the outboard side of the tokamak. HFS launch then both improves current drive efficiency which scales $\propto 1/N_\parallel^2$ \cite{Fisch1987} and enables off-axis LHCD in tokamak scenarios where it previously was not possible because waves launched at the low-field-side access limit would damp before reaching the top of an H-mode pedestal. Only one study verifying raytracing using full-wave simulation in reactor-relevant LHCD regimes has been performed previously \cite{Meneghini2010}, and a full-wave simulation of LH wave propagation and damping using HFS launchers has never been performed despite their importance in ARC reactor designs \cite{Sorbom2015}. Finally, in the DIII-D HFS cases the rays propagate through a number of caustics even though they undergo strong single-pass damping. This allows us to evaluate the effect of caustics on the predictive capabilities of raytracing in reactor relevant scenarios.

We simulated discharge \#174658 ($n_{e0}=4.5\times 10^{19}$ m$^{-3}$, $T_{e0}=6.75$ keV, $B_0 = 2.0$ T, and $N_\parallel = 2.7$) using numerical plasma profiles based on those measured in the experiment and equilibria obtained using EFIT. In TORLH the LHCD launcher was simulated using a waveguide placed 10 degrees below the inboard mid-plane launching an $N_\parallel = 2.7$. The TORLH waveguide was aligned and used the same $N_\parallel$ as the previous GENRAY simulations performed during the design of the DIII-D HFS launcher \cite{Seltzman2019}. The TORLH simulations performed here used $\mathcal{N}_m = 2047$ and $\mathcal{N}_\psi = 6000$ to ensure converged results. DIII-D had a larger than expected finite element requirement ((\ref{eq:feconv}) predicted only $\mathcal{N}_\psi \sim 3300$ would be needed in this case) resulting from rapid radial variations in the poloidal Jacobian used in the TORLH equilibrium representation induced by strong shaping and a large pedestal. Unless large numbers of finite elements were used in TORLH spectral pollution near the plasma separatrix dominated the solution. The $\mathcal{N}_m$ requirement in DIII-D was lower than the EAST cases performed in the next section despite the devices' similar physical dimensions. This is because the LH wave's $N_\parallel$ accessibility region in DIII-D is small relative to EAST and the waves are strongly damped before they experience substantial $N_\parallel$ variations. Our GENRAY simulations used 160 rays with a spectrum peaked at $N_\parallel = 2.7$. The ray spectrum  was slightly broadened with width $\Delta N_\parallel = 0.5$. The broadening of the GENRAY spectrum here was used to account for spectral broadening present within TORLH from the large waveguide (and possibly some small amount of diffraction).

Our simulations, shown in Figure~\ref{fig:diiid}, found excellent agreement between full-wave and raytracing simulations could be obtained in DIII-D. Previous raytracing studies in cases with strong single-passed damping have anticipated raytracing should be valid in reactor-like configurations. However, whether or not caustics would be a significant concern in raytracing codes in these cases was never firmly established. Here we demonstrate that despite a prominent caustic near the damping location raytracing accurately reproduces the full-wave results. These results indicate that standard raytracing techniques will indeed be sufficient in reactor-like LHCD scenarios, and it is unlikely that caustic surfaces will induce significant problems in raytracing simulations of reactors unless the caustic is precisely aligned with a flux surface power bin where there is substantial damping. While caustics may also induce some small amount of diffractional broadening not captured in raytracing this can be easily mitigated by slightly broadening the launched spectrum (spectral broadening from diffraction tends to symmetrically broaden the wave spectra \cite{Pereverzev1992}). This simulation of DIII-D is a key validation exercise for LHCD raytracing in integrated modeling for fusion reactors where present day full-wave models are far too computationally expensive to use, and improves our confidence in existing models of LHCD in reactor-relevant scenarios \cite{Najmabadi2006,Sorbom2015,Cardinali2017,Frank2020}.

\begin{figure}
    \centering
    \includegraphics[width=0.5\textwidth]{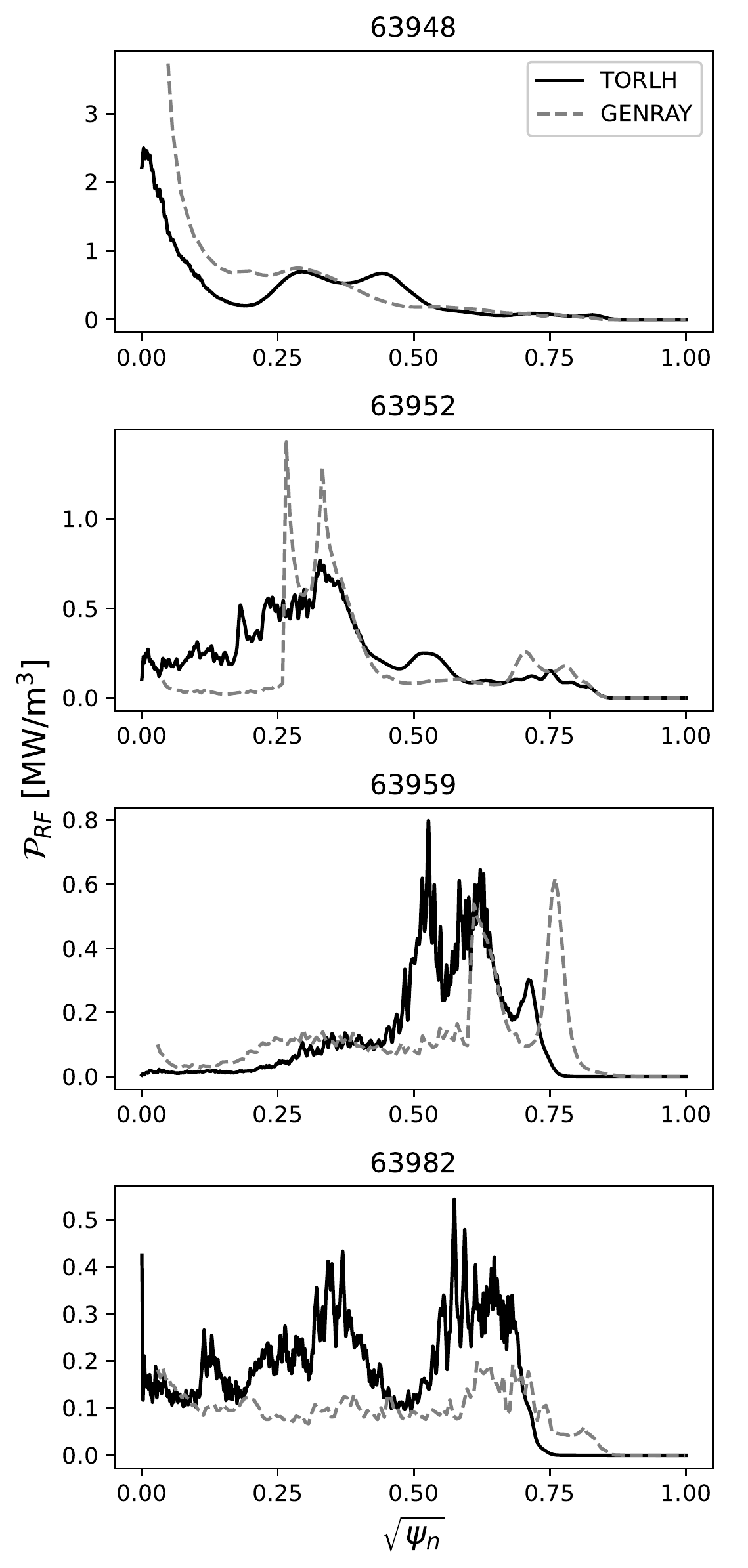}
    \caption{Power deposition profiles calculated by TORLH and GENRAY in EAST discharges 63948, 63952, 63959, and 63982. In the lowest density case, 63948, power deposition profiles in GENRAY and TORLH qualitatively agree. Slight differences in the peak locations are directly attributable to numerical discrepancies in the magnetic equilibrium reconstruction between TORLH and GENRAY. }
    \label{fig:east1}
\end{figure}

\begin{figure}
    \centering
    \includegraphics[width=0.5\textwidth]{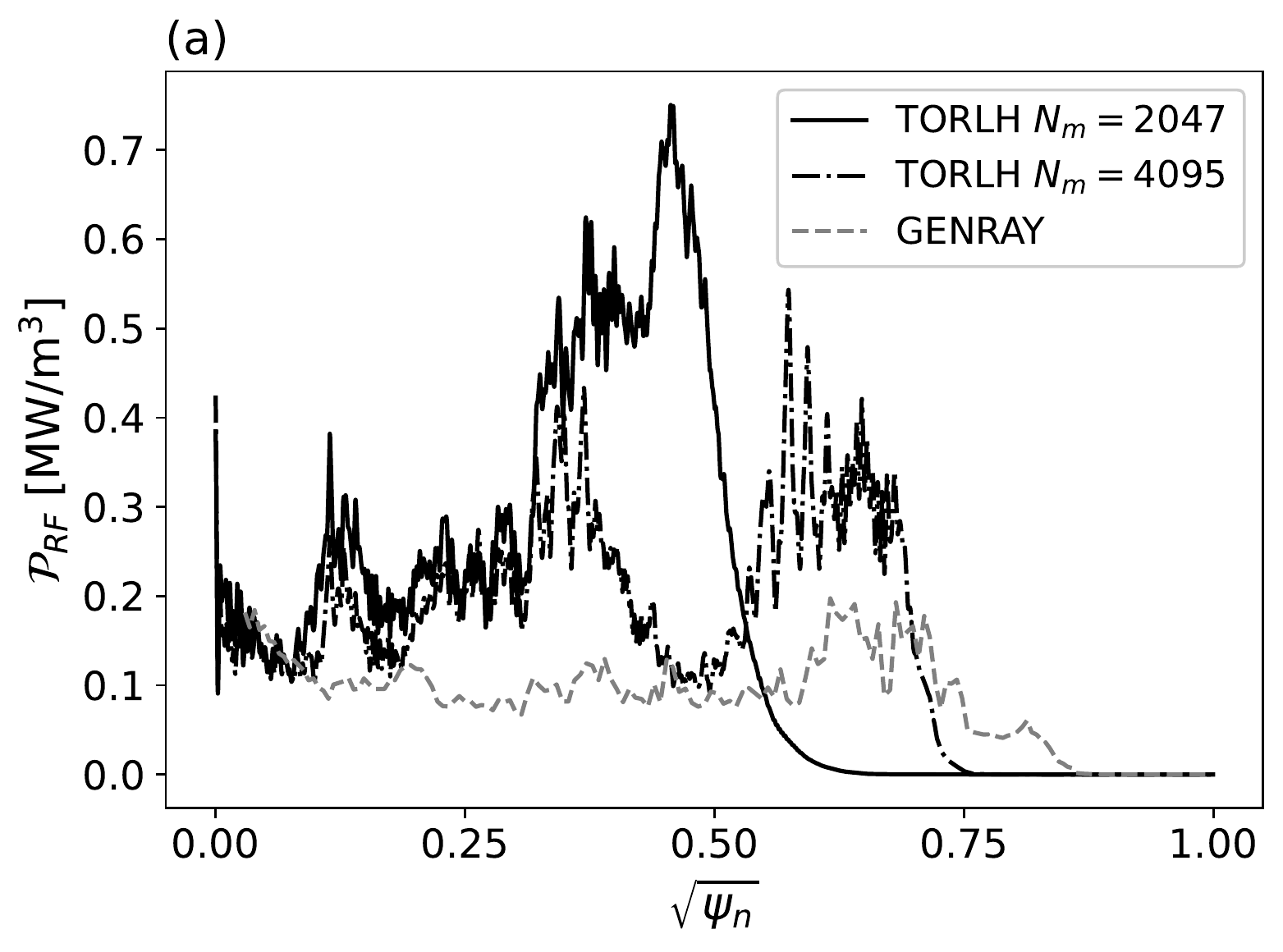}
    \includegraphics[width=0.5\textwidth]{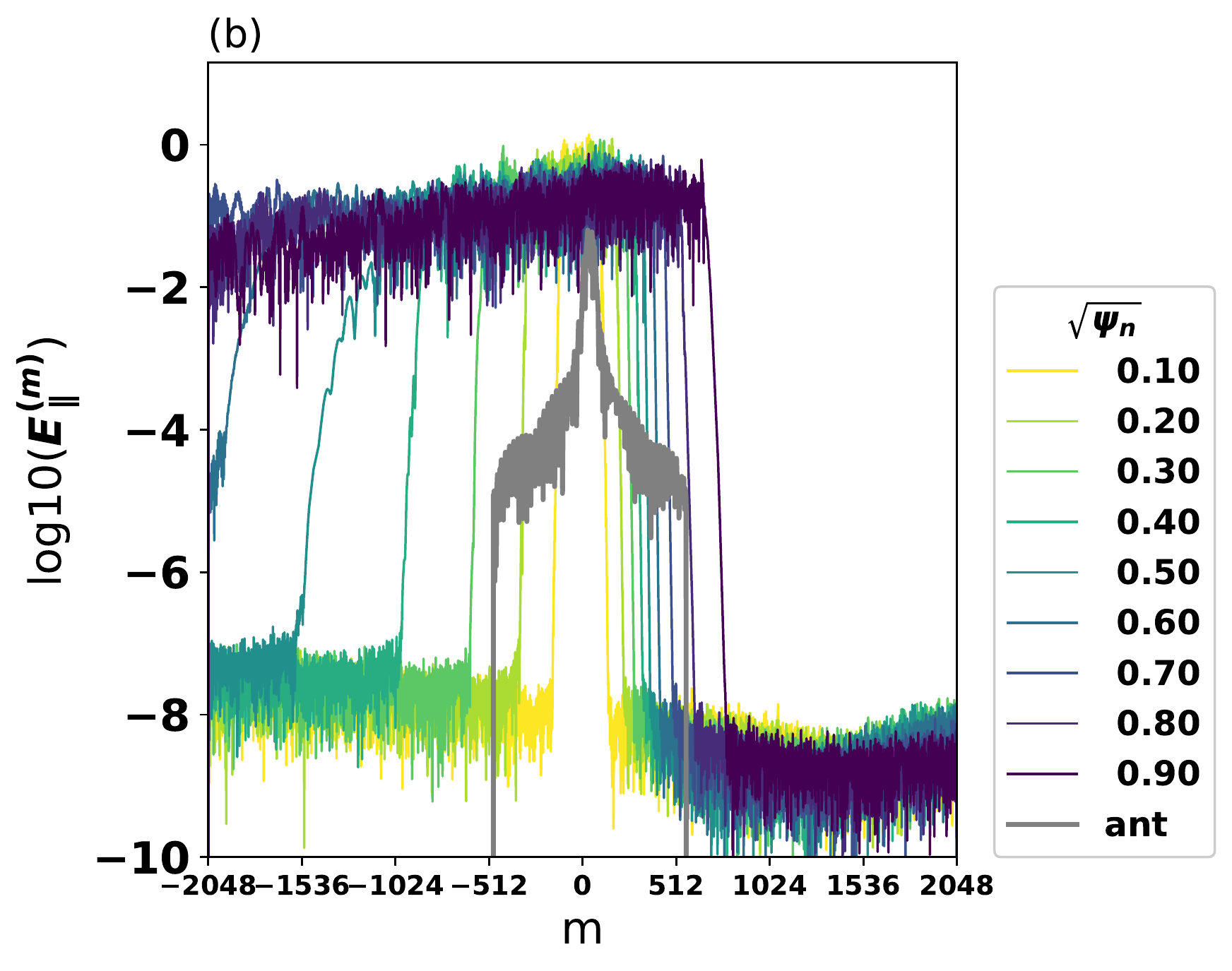}
    \caption{In the higher density EAST discharges 63959 and 63982 the TORLH simulations were not completely converged. Here, we show the convergence of the TORLH wave damping profile towards the raytracing power deposition profiles as the mode number is increased in 63982 (a) and the 69382 spectral convergence plot for $N_m = 4095$ (b).}
    \label{fig:east2}
\end{figure}  

\begin{figure}
    \centering
    \includegraphics[width=0.5\textwidth]{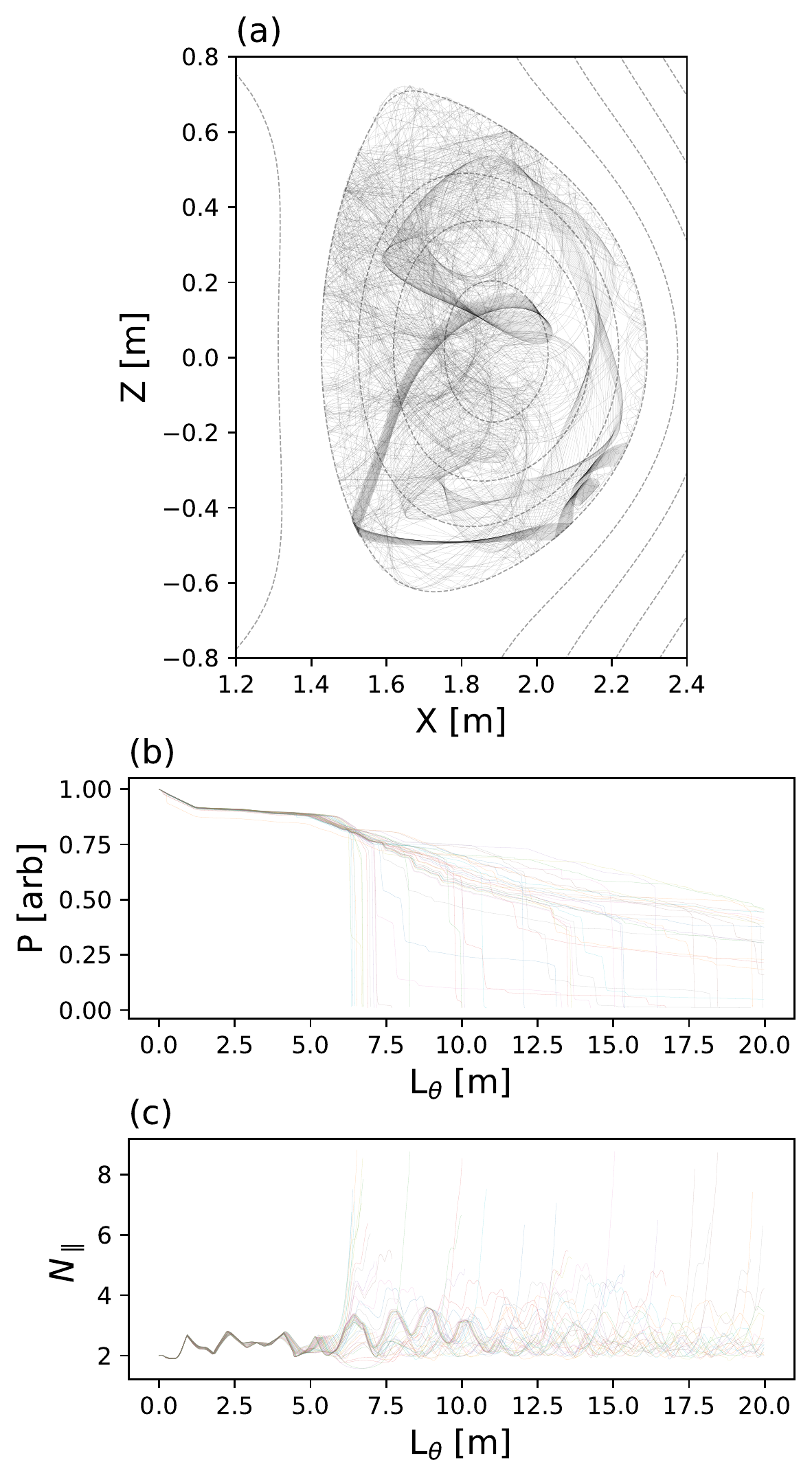}
    \caption{(a) The real-space poloidal plane trajectory of rays (rays stop when they damp $>95\%$ of their power in this figure), (b) the normalized ray power, and (c) the $N_\parallel$ evolution for a subset of rays in a simulation of EAST discharge \#63982. Many rays in this simulation propagate extremely long distances in poloidal arc length $L_\theta$, trapped between $N_\parallel \sim 2-3$, see (c), before suddenly depositing all of their power in the plasma as the result of a toroidally induced upshift shown in (b).}
    \label{fig:east3}
\end{figure}

\begin{figure*}
    \centering
    \includegraphics[width=0.5\textwidth]{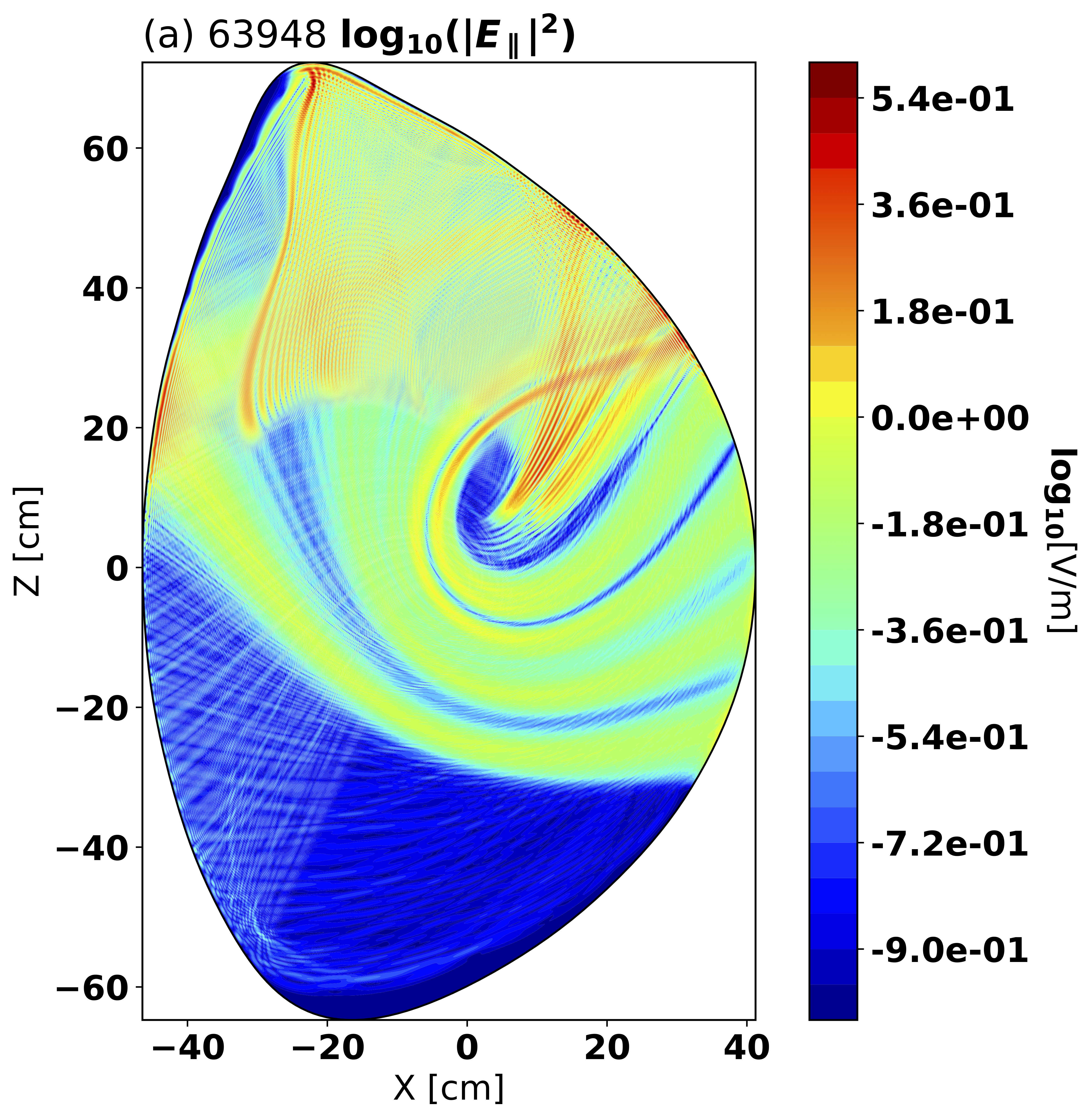}
    \includegraphics[width=0.48\textwidth]{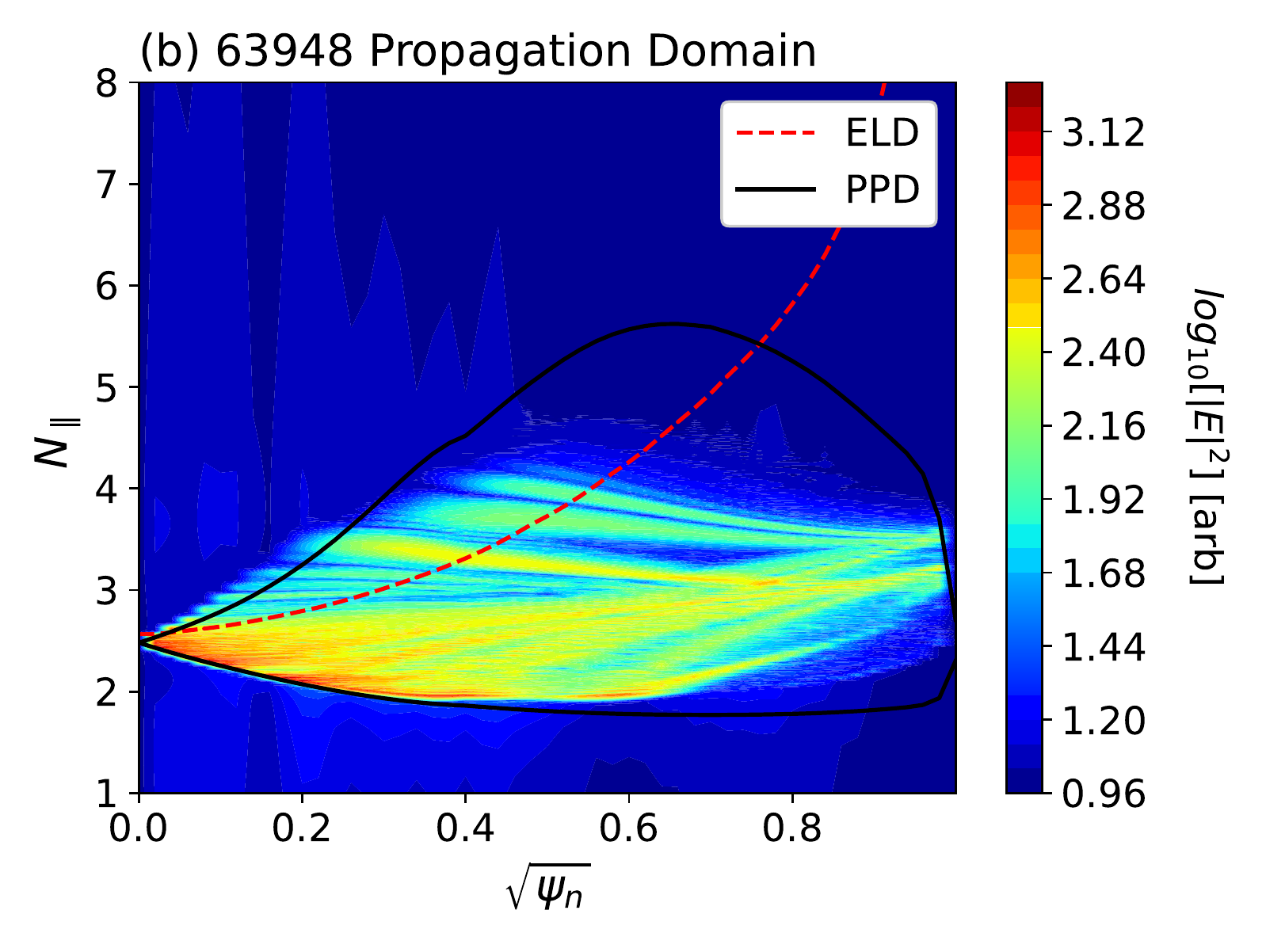}
    \includegraphics[width=0.5\textwidth]{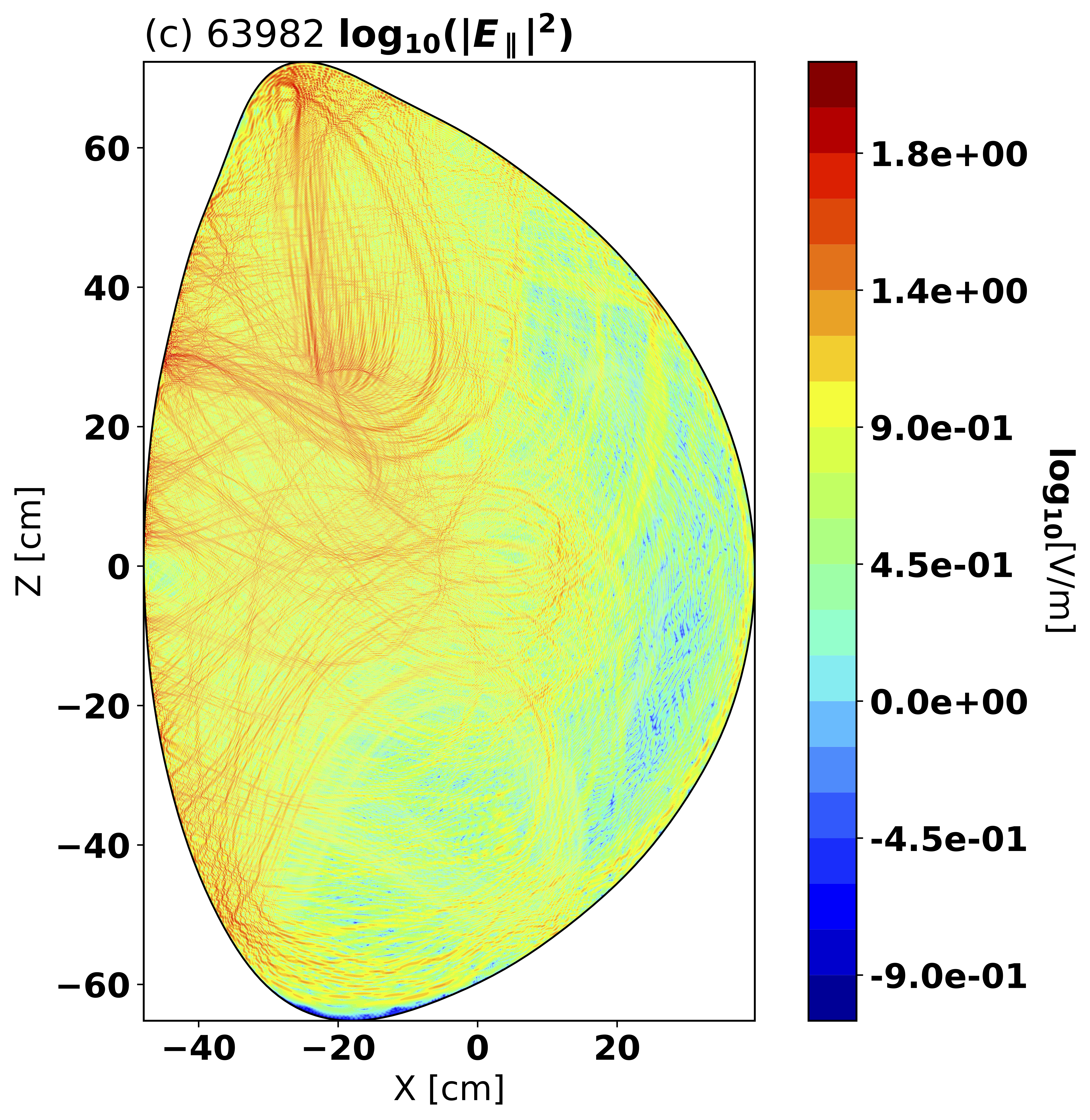}
    \includegraphics[width=0.48\textwidth]{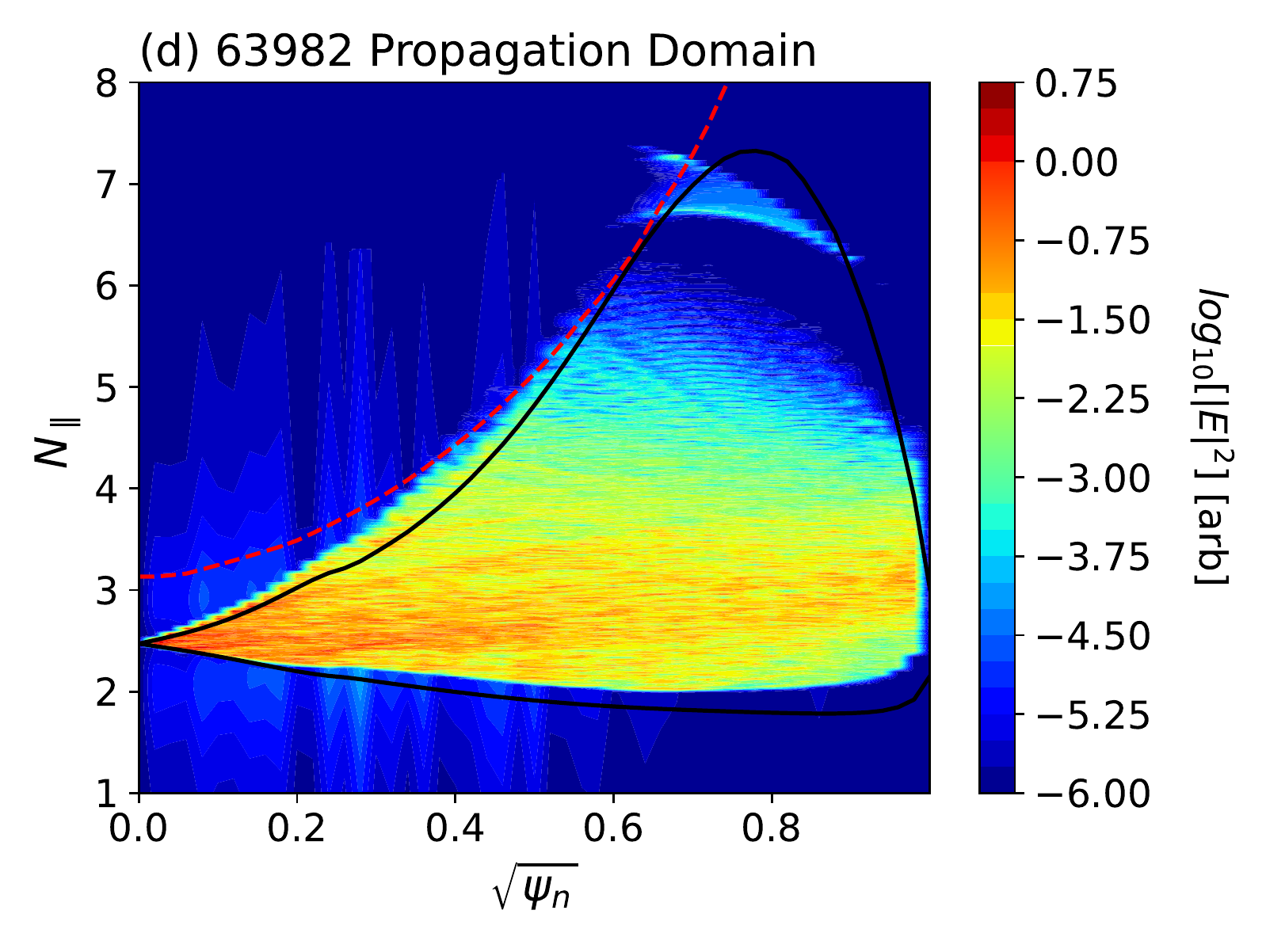}
    \caption{The $|E_\parallel|$ fields and power spectra overplotted on propagation domains for the strong-damping low-density shot, 63948, (a) and (b) and weak-damping high-density shot, 63982, (c) and (d) simulated with TORLH. The propagation domains in (b) and (d) are indicated by the solid black lines, and the $v_{ph\parallel} = 3 v_{the}$ limit corresponds to the dashed red line. The artifact at high $N_\parallel$ in figure (d) is a result of numerical noise from insufficient finite element resolution. Small deviations from the propagation domain are a result of the domains here not being exact (we have used the cold plasma electromagnetic propagation domains, not the hot plasma equivalent, as they are substantially easier to calculate).}
    \label{fig:east4}
\end{figure*}

\subsection{EAST Density Scan}\label{sec:sims_east}

The final set of raytracing and full-wave comparisons were performed using data from the EAST density scan experiments described in \cite{Garofalo2018}. In these experiments 4 non-inductive L-modes, shot numbers 63948, 63952, 63959, and 63982 where density is increasing with shot number, were generated.  These shots have been the subject of raytracing/FP simulations performed with both GENRAY/CQL3D and LUKE/C3PO, but these simulations were generally unable to reproduce the current profiles and integrated LH current obtained in the experiment. These simulations needed to use a modified launch spectrum and/or large anomalous diffusion coefficients to qualitatively replicate the experimental results \cite{Garofalo2018,Zhai2019}. 

One proposed explanation for the reduced predictive power of raytracing/FP simulations in this situation is the high aspect ratio in EAST where $\epsilon \sim 0.24$, substantially lower than C-Mod or DIII-D where $\epsilon \sim 0.33$. At high aspect ratio the toroidal upshift mechanism's capacity to close the spectral gap is significantly reduced \cite{Bonoli1981,Bonoli1982} leading to large components of the launched power spectrum becoming `trapped' in phase-space orbits. When trapped a wave will not experience a substantial toroidal $N_\parallel$ upshift causing it to damp until it has completed many passes through the plasma. In such situations other spectral gap closure mechanisms can become dominant, and this may explain the poor performance of conventional raytracing simulations. Here we attempt to evaluate the influence of `full-wave' gap closure from diffraction and interference on wave damping in EAST. GENRAY simulations used four 15 cm grills with 100 rays spread evenly over each placed at $\pm 12$ and $\pm 35$ degrees launching $N_{\parallel} = 2.0$ intended to emulate the EAST 4.6 GHz multi-junction launcher \cite{Liu2019}. The TORLH boundary condition was aligned to match the GENRAY launcher as precisely as possible and both simulations were initialized from identical plasma states in all cases. Performing TORLH simulations of EAST is extremely challenging as EAST is relatively large and the waves are weakly damped. In the TORLH simulations here $\mathcal{N}_m = 4095$, $\mathcal{N}_\psi = 3600$, and 32,000 CPU cores on Cori at NERSC were required for $\sim 1$ hr (for reference 32,000 core simulations entail using about 1/2 of the total Haswell CPU nodes on Cori). These were some of the largest, if not the largest, RF heating simulations ever performed, with problem sizes an order of magnitude greater than the previous record problem size in TORLH \cite{Yang2018}.

Power deposition profiles in EAST calculated with TORLH and GENRAY, shown in Figure~\ref{fig:east1}, generally agreed qualitatively at the lowest density. As density is increased LH wave damping becomes weaker and the power deposition profile flattens indicating weak damping from stochastic wave-fields. Although the same configuration procedure was used to achieve close matching between the raytracing and full-wave simulations as in Sections~\ref{sec:sims_cmod} and \ref{sec:sims_diiid}, EAST demonstrated larger discrepancies in the power deposition profiles. This discrepancy may be attributed to both two effects. The low resolution of the experimental EFIT, which in the more strongly damped cases, 63948 and 63952, caused noticeable error in the real-space wave trajectories due to the mismatch in magnetic field pitch angle between GENRAY and TORLH; and spectral convergence was inadequate in the higher density more weakly damped cases, 63959 and 63982. In these cases increasing the number of Fourier modes clearly provided better agreement with the raytracing solution. High $m$ components corresponding to high $N_\parallel$ off axis components of the wave power spectrum were resolved at higher $\mathcal{N}_m$ and the TORLH power deposition profile moved off axis like raytracing, a behavior seen in previous TORLH studies \cite{Yang2018} and demonstrated here in Figure~\ref{fig:east2}. Using convergence condition (\ref{eq:specconv1}) suggests $m\sim 4200$ are needed to resolve discharge 63982 and indicates the simulation is likely nearly converged but would benefit from greater resolution. However, it was not possible to increase $\mathcal{N}_m$ past 4095. For the TORLH solver to run effectively, $\mathcal{N}_m$ must have a value of $2^n - 1$ with integer $n$ \cite{Lee2014}. The computational resources required to perform $\mathcal{N}_m = 8191$ simulations do not exist as the memory requirement of TORLH solve scales with $\mathcal{N}_m^2$ and the simulations here, which used a large fraction of NERSC, were already memory limited.

Despite convergence difficulties and magnetic equilibria resolution limitations GENRAY and TORLH were demonstrated to have the same wave propagation dynamics. In the weakly damped simulations, 63958 and 63982, GENRAY and TORLH produce similar filamentary structures, shown in Figures~\ref{fig:east3}a and \ref{fig:east4}c. These structures result from the chaotic nature of LH wave propagation \cite{Bonoli1981,Bonoli1982,Bizarro1993}. A launched wave with even a slightly different initial spatial position will have much different late time behavior than other waves outside some small width where they remain relatively correlated. This leads to filamentary structures with width comparable to the initial correlation width. 

Another tool we can use to analyze wave propagation are cold plasma propagation domains (PPD). PPD plots were constructed using the methods from \cite{Paoletti1994,Zhai2019} and the TORLH power spectrum at the inboard midplane was obtained using the windowed Morlet-Gabor transform from \cite{DIppolito2003}. The PPD analysis, shown in Figure~\ref{fig:east4} demonstrates that TORLH is constrained by the same propagation domain constraints as raytracing has been shown to have in these EAST discharges \cite{Zhai2019}. Analysis of the power spectrum also shows GENRAY and TORLH  exhibit similar k-space trapping behavior indicating that full-wave effects, such as diffractional broadening, were not substantially contributing to spectral gap closure. In both GENRAY, see Figure~\ref{fig:east3}, and TORLH, see Figure~\ref{fig:east4}, spectral power became trapped at low $N_\parallel \sim 2-3$ then suddenly experienced a toroidal upshift and damped. This indicates that diffraction in TORLH, which would produce a more gradual transition to large $N_\parallel$ in the power spectrum, is not inducing sufficient broadening to detrap the low $N_\parallel$ component of the spectrum and induce spectral gap closure. Our analysis indicates both raytracing and full-wave simulations are dominated by spectral gap closure from the toroidal upshift mechanism and diffraction is not substantially modifying the gap closure process in the full-wave simulations. 

\section{Conclusion \& Discussion}

Using an upgraded version of the TORLH full-wave code we performed some of the largest simulations of RF-wave propagation in tokamaks to date providing high quality validation of Maxwellian LHCD raytracing simulations. The results of these simulations found that raytracing simulations could accurately replicate converged full-wave simulation results across a variety of different tokamaks. In these simulations diffraction and interference were not found to meaningfully contribute to spectral gap closure. There is some evidence that diffraction may slightly broaden the power deposition profiles, however, this effect generally seems to be small and differences in computational representations of the equilibrium, for example, are of comparable importance.

Our results in DIII-D provide confirmation of the validity of raytracing power deposition calculations in reactor-like configurations which include caustic surfaces. In C-Mod good agreement between raytracing and full-wave simulations was obtained in all cases including those that were weakly damped. The breakdown of the raytracing approximation at reflections and caustics did not result in significant differences in these simulations. This is perhaps not surprising as asymptotic analysis of the LH wave's reflection from cutoffs has shown that, except in the vicinity of the cutoff, the eikonal solution with a phase shift is recovered \cite{Richardson2010}. Quantitative agreement between raytracing and full-wave simulations in EAST was difficult to obtain, but propagation domain analysis did not indicate the presence of significant full-wave behaviour. This indicates the agreement in these simulations was limited by magnetic equilibrium resolution and convergence (fully converged simulations could not be performed with present computational resources). This leads us to conclude the influence of full-wave effects is generally small and spectral gap closure from diffraction and interference is unlikely. Current drive efficiency loss due to collisional damping of the LH wave in the scrapeoff layer \cite{Shiraiwa2011,Wallace2011} and spectral gap closure resultant from toroidal upshifts \cite{Bonoli1981,Bonoli1982}, scattering of LH waves from turbulence \cite{Decker2014,Biswas2020,Biswas2021,Baek2021} and parametric broadening of the LH wave spectrum \cite{Cesario2004} will likely be more important than full-wave effects. However, we must note that this revised TORLH analysis was not always converged and has so far been applied only to Maxwellian plasmas. Future work including converged simulations of large low aspect ratio tokamaks such as EAST and Fokker-Planck coupled simulations with non-Maxwellian wave damping that can provide accurate current drive calculations must be performed to entirely rule out full-wave effects. Additionally, implementation of a scrape-off layer model in TORLH with a method similar to that used by TORIC \cite{Shiraiwa_2017} would make it possible to rigorously study the effects of collisional damping and scattering from turbulence in the scrape-off layer using full-wave simulations and may be another avenue for future full-wave simulation work.

\section*{Acknowledgements}
This work was supported in part by Scientific Discovery Through Advanced Computing (SCIDAC) Contract No. DE-SC0018090 and Department of Energy grant: DE-FG02-91ER54109. This research used resources of the National Energy Research Scientific Computing Center, a DOE Office of Science User Facility supported by the Office of Science of the U.S. Department of Energy under Contract No. DE-AC02-05CH11231 using NERSC award FES-ERCAP0020035. The authors would like to thank S.J. Wukitch for providing the DIII-D equilibrium and profiles used in this study.

\section*{References}
\bibliography{docbib}

\end{document}